\documentclass[12pt]{article}
\usepackage{epsfig}

\newcommand{\be}{\begin{equation}}
\newcommand{\ee}{\end{equation}}

\def\bsg{\ifmmode B\to X_s\gamma\else $B\to X_s\gamma$\fi}
\def\bsll{\ifmmode B\to X_s\ell^+\ell^-\else $B\to X_s\ell^+\ell^-$\fi}
\def\shat{\ifmmode \hat{s}\else $\hat{s}$\fi}

\newcommand{\newc}{\newcommand}

\newc{\gsim}{\lower.7ex\hbox{$\;\stackrel{\textstyle>}{\sim}\;$}}
\newc{\lsim}{\lower.7ex\hbox{$\;\stackrel{\textstyle<}{\sim}\;$}}
\newc{\ie}{{\it i.e.}}
\newc{\etal}{{\it et al.}}
\newc{\mev}{\hbox{\rm\,MeV}}
\newc{\gev}{\hbox{\rm\,GeV}}
\newc{\tev}{\hbox{\rm\,TeV}}
\newc{\xpb}{\hbox{\rm\, pb}}
\newc{\xfb}{\hbox{\rm\, fb}}

\newc{\mtop}{m_t}
\newc{\mbot}{m_b}
\newc{\mz}{M_Z}
\newc{\mw}{M_W}
\newc{\alphasmz}{\alpha_s(M_Z)}
\newc{\swsq}{\sin^2\theta_W}
\newc{\cwsq}{\cos^2\theta_W}
\newc{\tw}{\tan\theta_W}
\newc{\cw}{\cos\theta_W}
\newc{\sw}{\sin\theta_W}
\newc{\BR}{\hbox{\rm BR}}
\newc{\zbb}{Z\to b\bar}
\newc{\Gb}{\Gamma (Z\to b\bar b)}
\newc{\Gh}{\Gamma (Z\to \hbox{\rm hadrons})}
\newc{\sgn}{\mbox{sgn}}

\newlength{\myem}
\newcounter{mysubequation}[equation]

\def\beq{\begin{equation}}
\def\eeq{\end{equation}}
\def\bea{\begin{eqnarray}}
\def\eea{\end{eqnarray}}
\def\slashchar#1{\setbox0=\hbox{$#1$}           
   \dimen0=\wd0                                 
   \setbox1=\hbox{/} \dimen1=\wd1               
   \ifdim\dimen0>\dimen1                        
      \rlap{\hbox to \dimen0{\hfil/\hfil}}      
      #1                                        
   \else                                        
      \rlap{\hbox to \dimen1{\hfil$#1$\hfil}}   
      /                                         
   \fi}                                         %
\catcode`@=11
\long\def\@caption#1[#2]#3{\par\addcontentsline{\csname
  ext@#1\endcsname}{#1}{\protect\numberline{\csname
  the#1\endcsname}{\ignorespaces #2}}\begingroup
    \small
    \@parboxrestore
    \@makecaption{\csname fnum@#1\endcsname}{\ignorespaces #3}\par
  \endgroup}
\catcode`@=12

\begin{document}

\baselineskip=18pt

\setcounter{footnote}{0}
\setcounter{figure}{0}
\setcounter{table}{0}

\begin{titlepage}
\begin{flushright}
HUTP-05/A0001 \\ SLAC-PUB-10928 \\ SU-ITP-04/44
\end{flushright}
\vspace{.3in}
\begin{center}
{\Large \bf  Predictive Landscapes and New Physics at a TeV }

\vspace{0.5cm}

{\bf N. Arkani-Hamed$^1$, S. Dimopoulos$^2$ and S. Kachru$^{2,3}$}

\vspace{.5cm}

{\it $^1$ Jefferson Laboratory of Physics, Harvard University,\\
Cambridge, Massachusetts 02138, USA}

{\it $^2$ Physics Department, Stanford University, \\ Stanford,
California 94305, USA}

{\it $^3$ SLAC, Stanford University, \\Menlo Park, California 94309, USA}

\end{center}
\vspace{.8cm}

\begin{abstract}
\medskip
We propose that the Standard Model is coupled to a sector with an
enormous landscape of vacua, where only the dimensionful
parameters---the vacuum energy and Higgs masses---are finely
``scanned" from one vacuum to another, while dimensionless couplings
are effectively fixed. This allows us to preserve achievements of
the usual unique-vacuum approach in relating dimensionless couplings
while also accounting for the success of the anthropic approach to
the cosmological constant problem. It can also explain the proximity
of the weak scale to the geometric mean of the Planck and vacuum
energy scales. We realize this idea with field theory landscapes
consisting of $N$ fields and $2^N$ vacua, where the fractional
variation of couplings is smaller than $\frac{1}{\sqrt{N}}$. These
lead to a variety of low-energy theories including the Standard
Model, the MSSM, and Split SUSY. This picture suggests sharp new
rules for model-building, providing the first framework in which to
simultaneously address the cosmological constant problem together
with the big and little hierarchy problems. Requiring the existence
of atoms can fix ratio of the QCD scale to the weak scale, thereby
providing a possible solution to the hierarchy problem as well as
related puzzles such as the $\mu$ and doublet-triplet splitting
problems. We also present new approaches to the hierarchy problem,
where the fine-tuning of the Higgs mass to exponentially small
scales is understood by even more basic environmental requirements
such as vacuum stability and the existence of baryons. These
theories predict new physics at the TeV scale, including a dark
matter candidate. The simplest theory has weak-scale ``Higgsinos" as
the only new particles charged under the Standard Model, with gauge
coupling unification near $10^{14}$ GeV.
\end{abstract}

\bigskip
\bigskip


\end{titlepage}

\section{A Predictive Neighborhood of the Landscape}

The most mysterious feature of the Standard Model is the extreme
smallness of its super-renormalizable couplings, the vacuum energy
$\Lambda$ and Higgs mass $m^2_h$, relative to the apparent cutoff of
the theory, the Planck mass $M_{{\rm Pl}}$. The tiny size of the
ratios $\Lambda^{1/4}/m_h \sim m_h/M_{{\rm Pl}} \sim 10^{-15}$ are
the essence of the cosmological constant and the hierarchy problems.
The naturalness hypothesis is that these ratios can be understood
dynamically without recourse to fine-tunings, and has been the
driving force for model-building in the last quarter-century. This
philosophy has been successfully applied to the hierarchy problem,
and led to theories of technicolor \cite{TC}, the supersymmetric
standard model (SSM) \cite{dg}, large extra dimensions \cite{add},
warped compactifications \cite{rs} and little Higgs theories
\cite{lh}, as possible natural approaches to the hierarchy problem,
all with experimentally testable consequences at the LHC. In
contrast, there is not a single natural solution to the cosmological
constant problem. Still, until the late nineties there was a hope
that some hidden symmetry of string theory might set the energy of
the vacuum to zero. It seemed inevitable that a dynamical mechanism
that could reduce the vacuum energy by 120 orders of magnitude would
not just stop there, but take it all the way down to zero. This hope
was severely challenged in the late nineties, with the discovery of
the accelerating universe and indication of a non-zero vacuum
energy. The cosmological constant problem became suddenly harder, as
one could no longer hope for a deep symmetry setting it to zero.

Yet, already in the eighties, an argument due to Weinberg
\cite{Weinberg} had anticipated the correct order of magnitude of
the vacuum energy. It was based on the  ``multiple vacua" or
``multiverse'' hypothesis \cite{anthropicrefs} which states that
there is an enormous number of vacua (or universes) all with
identical physics---spectra, forces and parameters--- as the
standard model. Assuming that they differ from each other
\textit{only} in the value of the vacuum energy, the existence of
gravitationally clumped structure in the form of galaxies excludes
all universes with vacuum energy significantly larger than ours and
led to the prediction of a vacuum energy of the correct order of
magnitude \cite{Weinberg}. This hypothesis has recently been gaining
momentum because of gathering theoretical evidence that there is a
vast ``landscape'' of vacua in string theory
\cite{BP,GKP,MSS,KKLT,Lenny,Douglas}, each with different low-energy
theories and parameters.

The landscape suggests an enormous epistemological shift in particle
physics, possibly analogous to that caused by the realization that
there is not just one, but a large number of solar systems in the
universe, and our planetary distances are just historic accidents of
no fundamental importance. It points to the possibility that our
laws of nature---the particle spectrum, forces and magnitudes of the
parameters of the standard model---may, at least in part, also be
accidents of the vacuum that we happened to land in, and the age-old
quest of trying to understand them further is unlikely to be
intimately related to the underlying fundamental theory.

The landscape, on the one hand, offers new tools for addressing old
problems, for making predictions, and for correlating observables.
One new set of tools are environmental (or anthropic) vacuum
selection rules, such as Weinberg's ``galactic principle'' (or
structure principle). Another are statistical arguments
\cite{Douglas,Dvali,Ashok, DD,
GKT,CQ,DGKT,KW,Dvalitwo,Lust,DDtwo}---looking for features or
correlations that are statistically favored in the landscape. On the
other hand, the landscape leads us to question the relevance of the
questions we have been asking (by drawing parallels to planetary
distances) and the validity of the traditional methodology for
making predictions and finding correlations, based on symmetries and
dynamics. This methodology, however, has had some remarkable,
hard-to-dismiss successes in relating renormalizable
\textit{dimensionless} quantities, such as gauge or Yukawa
couplings. An example is gauge coupling unification in SUSY theories
\cite{dg,drw}. These successes suggest that, at least for some
parameters, the traditional methods should be applicable, in spite
of the presence of the landscape. In fact, even Weinberg's original
prediction of the vacuum energy relies on considering all parameters
of the standard model as fixed, and varying \textit{only} the vacuum
energy. His prediction would have been much weaker if several
parameters, such as $n_B/n_\gamma, \delta \rho/\rho, \alpha$, etc.,
were allowed to vary (see e.g. \cite{Tegtwo, Graesser} and the
discussion in the next subsection). So, to justify Weinberg's
argument, it is essential to find a rationale for why only the
vacuum energy is being scanned in the landscape.

The success of Weinberg's argument as well as the successes of the
usual methodology for particle physics can \textit{both} be
preserved provided we live in a ``predictive'' (or ``friendly'')
neighborhood of the landscape, where only the super-renormalizable,
dimensionful parameters of the theory--such as the vacuum energy and
the Higgs mass in the Standard Model--are finely scanned. These are
the very quantities that are not protected by symmetries and are
associated with the fine-tuning problems associated with the
electroweak hierarchy and the cosmological constant. Though this may
seem like a lot to hope for, in section 2 we will present simple
examples of field-theoretic landscapes accomplishing this. In the
remainder of this section, we use this tool to address some
important puzzles related to the CC and hierarchy problems.

\subsection{The Mystery of Equidistant Scales}

This refers to the geometric relation between the weak, Planck and
CC scales:

\begin{equation}
\Lambda^{1/4} \sim \frac{v^2}{M_{Pl}}
\end{equation}

It is tempting to speculate that this signals a deep, unknown,
\textit{dynamical} connection between the three most important
masses in physics. Instead, we will now argue that in a wide class
of theories, as long as we are in a predictive neighborhood, this
relation follows from nothing more than the environmental
requirement of the existence of galaxies---a generalization of
Weinberg's argument.

Recall that the essence of Weinberg's argument is that the
cosmological constant $\Lambda$ must be smaller than the energy
density of the universe when galaxies (non-linear structures) start
to form. Moreover, structure first starts growing when the universe
becomes matter dominated, and after that point $\delta \rho /\rho$
grows like the scale factor $a$, so expansion by an additional
$(\delta \rho/\rho)^{-1}$ is necessary before non-linear structures
form. Therefore, Weinberg's argument bounds
\begin{equation}
\Lambda \lsim \rho_{MR} \left(\frac{\delta \rho}{\rho}\right)^3
\end{equation}
where $\rho_{MR}$ is the energy density of the universe at
matter-radiation equality.

Furthermore, in theories where the matter density is dominated by
weakly interacting dark matter particles of mass $m_{DM}$, the
standard perturbative freezout calculation gives us
\begin{equation}
\rho_{MR} \sim \left(\frac{m_{DM}^2}{\alpha^2 M_{Pl}}\right)^4
\end{equation}
where $\alpha$ is a weak-coupling factor. Weinberg's bound then
reads
\begin{equation}
\Lambda^{1/4} \lsim \frac{(\frac{\delta
\rho}{\rho})^{3/4}}{\alpha^2} \frac{m_{DM}^2}{M_{Pl}}
\end{equation}

In the absence of any other reason for small $\Lambda^{1/4}$, the
upper limit should be roughly saturated. So, in any theory where the
DM particle is naturally at the weak scale $v$---such as the
SSM---we parametrically predict the relation (1). Furthermore, in
inflationary theories where $\delta \rho/\rho$ is determined by the
same weak-coupling factors that set $\alpha$, it is reasonable to
expect a rough cancellation between the numerator and denominator,
and if $(\frac{\delta \rho}{\rho})^{3/4}$ is parametrically $\sim
\alpha^2$, relation (1) follows. To maintain this relation it is
crucial to be in a predictive neighborhood. In a more general place
in the landscape where many quantities are scanned, $\frac{\delta
\rho}{\rho}$ and $\alpha$ can vary, the relation $(\frac{\delta
\rho}{\rho})^{3/4} \sim \alpha^2$ will be violated, and the
numerical prediction (1) will be lost. Similarly, at a general place
in the landscape both $\Lambda$ and $m_{DM}$ are variable and
equation 4 allows galaxy formation in universes with large values of
$\Lambda$ and $m_{DM}$. It would then be puzzling why such large
values are not realized in nature since they seem favored, as they
reduce the CC tuning. In fact $\Lambda$ and $m_{DM}$ could be bigger
by a factor of about $10^4$  (thereby reducing the CC tuning by
$10^{16}$) and still allow for star formation in a galaxy \cite{Z}.
So, to avoid this puzzle, it is crucial to be in a neighborhood
where $m_{DM}$ is fixed. In theories where  $m_{DM} \sim v$ this can
happen because of the ``atomic principle" to which we turn in the
next subsection.

Note that in order to get interesting structure like galaxies, it is
not sufficient to have $\delta \rho/\rho$ grow to be of ${\cal
O}(1)$. While this would allow gravitational clumping for Dark
Matter, this just forms large virialized DM balls, which do not
further fragment. One needs baryons to further cool and clump to
form smaller scale structures.

This more detailed version of the structure principle can be used to
answer the question ``why is $m_h^2 < 0$?''. In the SM, there is no
phase transition as $m_h^2$ passes through zero. Even for $m_h^2
> 0$, electroweak symmetry is still broken down to $U(1)_{EM}$ by
quark condensation in QCD.  Meanwhile, the Yukawa couplings still
break the light fermion chiral symmetries, and indeed integrating
out the heavy Higgs induces four-fermion operators that turn into
fermion masses after the quark condensates form.   So, why did
Nature choose to break EW symmetry with a Higgs vev, rather than by
QCD itself? A possible answer is that, in the worlds where the
electroweak symmetry is broken by QCD (see \cite{nucl} for a nice
discussion of these worlds), any baryon asymmetry present in the
early universe is wiped out after the QCD phase transition. As
discussed in greater detail in an appendix, this is because baryon
number violation from the weak interactions very effectively
destroys any existing baryon asymmetry. Electroweak interactions
violate $B + L$ via the anomaly. At temperatures above the weak
phase transition, therefore, a non-zero $B-L$ is roughly equally
divided between the quarks and leptons. In our universe, at
sufficiently low temperatures below the EW phase transition but
above the QCD pase transition, the $B$ violation shuts off, and we
are left with a locked-in net $B$ number, which is converted to
nucleons after the QCD phase transition. The situation is very
different in the case where QCD itself breaks the EW symmetry. Since
the baryons are {\it heavier} than the W, the B violating
interactions remain in equilibrium at temperatures beneath the
nucleon masses, and therefore the non-zero $B-L$ is almost all in
leptons. In fact, the baryon number is suppressed down to its
freezeout value in a baryon-symmetric universe.

\subsection{The Atomic Principle}

In a friendly neighborhood  only the super-renormalizable
parameters $\Lambda$
and $m_h^2$ (and therefore the weak vev $v$) are finely scanned,
while all other parameters are approximately constant. This is the
case, in particular, for all the other dimensionful
parameters---such as $\Lambda_{QCD}$---that are determined by
dimensional transmutation and thus are naturally much smaller that
$M_{Pl}$. Here we briefly review how environmental arguments peg $v$
to $\Lambda_{QCD}$, and can therefore account for the smallness of
$v$.

The strategy, just as in the case of the CC, is to focus on some
crucial infrared property of the universe which sensitively depends
on $v \over \Lambda_{QCD}$, in this case the existence of atoms
\cite{donoghue} (``atomic principle''). In a predictive
neighborhood the Yukawa couplings that determine the quark and
lepton masses are fixed, so as we vary $v$ all these masses scale
simply with $v$. As $v$ increases the neutron-proton mass difference
increases, till eventually nuclei heavier than hydrogen, all of
which contain neutrons, become unstable. Conversely, as $v$
decreases the neutron eventually becomes lighter than the proton and
the hydrogen atom disappears. So the existence of both hydrogen and
heavier nuclei constrains $v$ to be within a factor of a few of its
measured value. Note that in a general neighborhood of the landscape
the possibility that changes in $v$ could be compensated by
corresponding changes in the Yukawa couplings would make it
impossible to precisely determine $v$ from the atomic principle. So,
only in a friendly neighborhood can we understand the proximity of
the weak and the QCD scales.

\subsection{The Landscape and the Little Hierarchy problem}
In the over two decades of work on the hierarchy problem, we have
faced a persistent difficulty. On the one hand, naturalness suggests
that there should be a new set of particles at the weak scale to
solve the hierarchy problem. On the other hand, indirect bounds on
higher dimension operators in the Standard Model are typically in
excess of the TeV scale, from the GUT scale for baryon number
violating operators to the $\sim 100-1000$ TeV scale for flavor and
CP violating operators, to $\sim 5 - 10$ TeV for flavor-conserving
dimensions six operators contributing to precision electroweak
observables. In other words, the new particles needed for
naturalness at the TeV scale have had ample opportunities to reveal
themselves indirectly in various processes, from precision
electroweak observables, to $K-\bar{K}$ mixing, $B$ meson mixing and
decays like $B \rightarrow s \gamma$, $\mu \to e \gamma$, as well as
electron and neutron dipole moments, and yet have not manifested. As
is well-known, technicolor faces it most serious challenges from
precision electroweak tests as well as excessive flavor-changing
neutral currents. SUSY also suffers from a flavor problem, and one
might have also expected non-zero electric dipole moments and
non-standard $B$ physics, together with a light Higgs which has yet
to materialize, making the MSSM already tuned to the few percent
level. This tension between the requirements of naturalness on the
one hand and the absence of indirect evidence for new TeV scale
physics on the other hand is known as the ``little" hierarchy
problem \cite{barbieri}.

It is difficult to know how seriously to take this problem--after
all it is not as dramatic a difficulty as the hierarchy problem
itself. Indeed, almost all the work in physics beyond the standard
model has revolved around modifications of the minimal models that
address these nagging problems. But one can easily imagine that
things could have turned out differently. Had the $S$ parameter been
measured experimentally to be, say, $S = 2$, most model-builders
wouldn't be thinking about what to make of the landscape--they would
be trying to find the correct underlying technicolor model of the
weak scale. Similarly, if the Higgs was discovered with a mass
beneath the $Z$ mass, many people would be firmly convinced of
weak-scale SUSY, even more so if, say, deviations in $b \rightarrow
s \gamma$ and an electron EDM had also been discovered. Instead we
are left with the puzzle of why no clear deviations from the
Standard Model have yet shown up. It could be that there is just a
percentish fine-tuning in the underlying theory so that the new
particles are a little heavier than we expected. Or, there may be a
mechanism at work ``hiding" the new particles in a natural theory of
the weak scale from indirect processes, as happens e.g. in little
Higgs models with T-parity \cite{Tparity} or in supersymmetric
models that allow for a heavier Higgs \cite{fat}, although all such
theories require a certain amount of model-building engineering to
work.

Landscape reasoning suggests a very different possibility. The
little hierarchy problem could be telling us something of great
structural importance about weak scale physics. If environmental
arguments, such as the atomic principle, can explain the value of
the weak scale, there is no need for a large number of new particles
and interactions beyond the standard model, and therefore the
success of the Standard Model extrapolated to energies well above
the weak scale is simply understood.

\subsection{Split SUSY}

The simplest expectation for the low energy theory emanating from a
friendly neighborhood, together with the atomic and structure
principles, is just the SM. This would entail giving up the two
successes of the SSM, unification and DM. Fortunately, this is not
necessary. Some well motivated extensions of the SM contain new
fermions which can be protected by approximate chiral symmetries,
ensuring their presence at low energies.  The simplest such
possibilities are SUSY extensions of the SM, with SUSY broken at
some high scale, but SUSY-fermions (gauginos and Higgsinos)
protected by an R-type symmetry \cite{AD,Split,Splittwo,AntD}. These
Split-SUSY theories preserve the successes of the SSM and, most
important, naturally account for the problematic absence of SUSY
signatures, such as light Higgs and sparticles, proton decay, FCNC
and CP-violation. They are minimal enlargements of the SM with small
and just-right particle content to account for unification and DM.
Motivated by the landscape, in this paper we will propose other such
extensions of the SM which share this feature of minimality and have
distinct experimental signatures.

\subsection{Living Dangerously on the Landscape}

In addition to anthropic arguments, such as the structure or atomic
principles, it is a priori possible to get further information by
the use of statistical arguments. In practice these require a
thorough understanding of both the landscape and the way in which
the various vacua get populated in cosmological history---a
difficult task at best. Fortunately, such a detailed understanding
is often unnecessary. This happens when the anthropically allowed
range of the theory is so narrow that the statistical distribution
of the vacua should be flat there \cite{Weinberg}. Then the expected
value of the corresponding physical quantity should be at a typical
point of the allowed range. For the CC this would be of order of the
maximal value allowed by the structure principle. The general
corollary is that anthropic reasoning leads to the conclusion that
we live dangerously close to the edge of violating an important but
fragile feature of the low-energy world---such as the existence of
galaxies or atoms. This will be a recurring theme in the
landscape-motivated models that we will propose here.

\section{Field-theory landscapes}

The idea that there may be a vast landscape of metastable vacua
leading to an enormous diversity of possible low-energy environments
has recently gained momentum in string theory. However as yet, the
vacuum leading to the Standard Model at low energies has not
emerged, so it is difficult to infer the consequences of this
landscape for the structure of physics beyond the Standard Model.

Because of this, we wish to instead explore an effective field
theory description of a landscape of vacua, which will allow us to
map out the rules for landscape model-building as well as the broad
consequences for particle physics in a concrete way. As is
well-known, the existence of $\sim 10^{500}$ vacua in string theory
is not directly the consequence of any detailed ultraviolet property
of quantum gravity, but follows simply from a large number $N$ of
fluxes and moduli, with the number of vacua being exponentially
large in $N$. This property can be very simply reproduced in
effective field theories. We will consider explicit examples of such
field theories, which will allow to study properties of the
associated landscape of vacua and the couplings to the Standard
Model in detail. Toy field theory landscapes have also been recently
discussed in \cite{DDG}.

\subsection{Non-supersymmetric  landscape}

We will begin by considering theories which are non-supersymmetric
all the way up to the UV cutoff $M_*$ of the effective theory,
though of course the deep UV theory of quantum gravity may well be
supersymmetic. We will turn to supersymmetric theories in the next
subsection; as we will see, SUSY introduces special advantages over
completely non-SUSY theories.

Consider a single scalar field $\phi$ with a general, quartic
potential, including both cubic and linear terms in $\phi$ (so there
is no $\phi \to - \phi$ symmetry). We will assume that all the mass
parameters are $\sim M_*$ (though of course formally, they must be
parametrically smaller than $M_*$ in order to be able to trust the
low-energy effective theory for $\phi$). If the mass squared
$m_\phi^2 < 0$, the theory has two minima with $\langle \phi \rangle
= \phi_{\pm}$, with vacuum energies $V_{\pm}$, where we take $V_+
\ge V_-$. The false vacuum can be exponentially long-lived; in the
thin-wall approximation, the tunneling rate per unit volume is
\begin{equation}
\Gamma \sim M_*^4 e^{-27 \pi^2 \frac{\sigma^4}{p^3}}
\end{equation}
where $\sigma$ is the surface tension of the bubble (set by the
barrier height) and $p$ is the pressure in the bubble (set by the
difference is vacuum energies). Even if $p^{1/4} \sim \frac{1}{2}
\sigma^{1/3}$, (corresponding e.g. to a cubic term that is $\sim 1/2
M_*$), for $M_*$ close to the GUT scale, the false vacuum is stable
on cosmological time scales. We will label the vacua by $\eta = \pm
1$, and define
\begin{equation}
V_\eta = V_{\rm{av}} + \eta V_{\rm{dif}}, \, \, {\rm where} \, \,
 V_{{\rm av}} = \frac{1}{2}(V_+ + V_-), \, V_{{\rm dif}} =
\frac{1}{2}(V_+ - V_-)
\end{equation}

\begin{figure}
\begin{center}
\epsfig{file=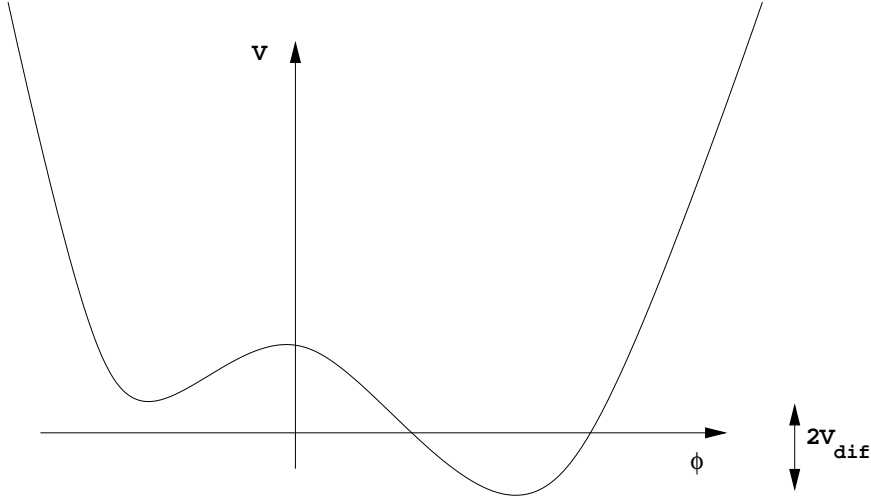,width=0.7\textwidth}
\end{center}
\caption{The potential $V(\phi)$.}
\end{figure}

Now, consider $N$ such sectors, with scalars $\phi_i$ for $i = 1,
\cdots, N$, and independent quartic potentials $V_i(\phi_i)$, so
that
\begin{equation}
V = \sum_{i = 1}^N V_i(\phi_i)
\end{equation}
This theory has $2^N$ vacua, labeled by $\{\eta\} = \{\eta_1,
\cdots, \eta_N\}$, with $\eta_i = \pm 1$. We will assume that $N
\sim 10^2 - 10^3$, so that we have the $\sim 10^{120}$ vacua needed
for Weinberg's resolution of the cosmological constant problem. This
is our exponentially large ``landscape" of vacua.

Note that we have assumed that there are no cross-couplings between
the different scalars $\phi_i$. Our results are unaffected as long
as such couplings can be treated as perturbations. This form of the
potential might arise, for instance, if the different $\phi_i$ are
stuck to living on different points in extra dimensions. The
requirement that these points be separated by more than the UV
cutoff $M_*$ leads to a mild constraint on the size $R$ of $n$ extra
dimensions $R \gsim M_*^{-1} N^{1/n}$. We should, however, examine
the cross-couplings that are inevitably induced by gravity, which of
course does couple to all the $\phi_i$ with equal strength. To begin
with, note that the presence of a large number $N$ of fields makes
gravity parametrically weaker, as a loop of these fields generates a
quadratically divergent correction to the Einstein-Hilbert term in
the effective action, so that
\begin{equation}
M_{{\rm Pl}}^2 \gsim \frac{N}{16 \pi^2} M_*^2
\end{equation}
Now, a 1-loop diagram does induce cross-couplings between the
different $\phi_i$, for instance generating a term of the form
\begin{equation}
\frac{1}{16 \pi^2} \sum_{i,j} \frac{V_i(\phi_i) V_j(\phi_j)}{M_{{\rm
Pl}}^4}
\end{equation}
but from our above bound on $M_{{\rm Pl}}^2$, this correction is
smaller than
\begin{equation}
\frac{16 \pi^2}{N} \sum_i V_i(\phi_i) \left(\frac{1}{N} \sum_j
V(\phi_j) \right)
\end{equation}
which for large $N$ is subdominant to the leading $\sum_i
V_i(\phi_i)$ term we started with. Thus, even including the
universal interactions with gravity, our simple form of the
potential, being the sum of $N$ independent potentials, is
consistent and radiatively stable.

\subsubsection{Statistical Interlude}

Given these $2^N$ vacua, we would like to know how various
parameters of the theory vary or ``scan" as we go from one vacuum to
another. Let us begin by asking this question for the vacuum energy.
The vacuum energy $\Lambda_{\{\eta\}}$ of the vacuum $\{\eta_1,
\cdots, \eta_N \}$ is given by
\begin{equation}
\Lambda_{\{\eta\}} = \bar{\Lambda} + \sum_i \eta_i V_{{\rm dif} \,
i}
\end{equation}
where
\begin{equation}
\bar{\Lambda} = \sum_i V_{{\rm av} \, i} = N \bar{V}_{{\rm av}}
\end{equation}
Let us define
\begin{equation}
\rho(\Lambda) = \sum_{\{\eta\}} \delta(\Lambda - \Lambda_{\{\eta\}})
\end{equation}
to be the distribution of vacuum energies in our theory, so that the
number of vacua ${\cal N}(\Lambda_1,\Lambda_2)$ with energy between
$\Lambda_1$ and $\Lambda_2$ is given by
\begin{equation}
{\cal N}(\Lambda_1,\Lambda_2) = \int_{\Lambda_1}^{\Lambda_2} d
\Lambda \rho(\Lambda)
\end{equation}

Obviously, we could determine $\rho(\Lambda)$ or ${\cal
N}(\Lambda_1,\Lambda_2)$ directly by examining all $2^N$ vacua and
making a histogram of all the vacuum energies. But at large $N$,
standard statistical arguments, familiar from the derivation of the
central limit theorem (which we review in an appendix), tells us
that $\rho(\Lambda)$ is well-approximated by a Gaussian. In general,
given a quantity
\begin{equation}
F_{\{\eta\}} = \sum_{i = 1}^N \eta_i f_i
\end{equation}
the corresponding distribution
\begin{equation}
\rho(F) = \sum_{\{\eta\}} \delta(F - F_{\{\eta\}})
\end{equation}
becomes well-approximated by a Gaussian
\begin{equation}
\rho(F) \to \frac{2^N}{\sqrt{2 \pi N f^2}} \, e^{-\frac{F^2}{2N
f^2}}
\end{equation}
where
\begin{equation}
f^2 = \frac{1}{N} \sum_i f_i^2
\end{equation}
and corrections of order ${\cal O}(1/N^2)$ in the exponential. Note
the familiar $\sqrt{N}$ factor in the width of the Gaussian. Note
also that, for $F \ll {\sqrt N} f$, the distribution is
approximately flat, and that the number of vacua between $(-F, F)$
is
\begin{equation}
{\cal N}(-F,F) \to \frac{2^{N+1}}{\sqrt{2 \pi N}} \frac{F}{f}
\end{equation}
so that the vacuum with smallest $F$ will have
\begin{equation}
F_{\min} \sim \frac{\sqrt{2 \pi N} f}{2^N}
\end{equation}

As an example, we chose $N = 15$, and picked 15 random values for
$f_i$ in the range $(0,1)$. For these $2^{15} = 32768$ vacua, we
numerically determined ${\cal N}(-F,F)$,  and compared with the
statistical expectation. The result is shown in fig. 1 (for small
$F$); the bumpy line is the actual distribution, while the straight
line is the statistical prediction. The agreement is excellent, as
to be expected, since the corrections in the exponent are $\sim
1/N^2$, less than one percent. We also show a close-up of the
distribution for the smallest values of $F$ in Fig. 2, confirming
that the tiniest values are indeed of order $\sim 2^{-N}$.

\begin{figure}
\begin{center}
\epsfig{file=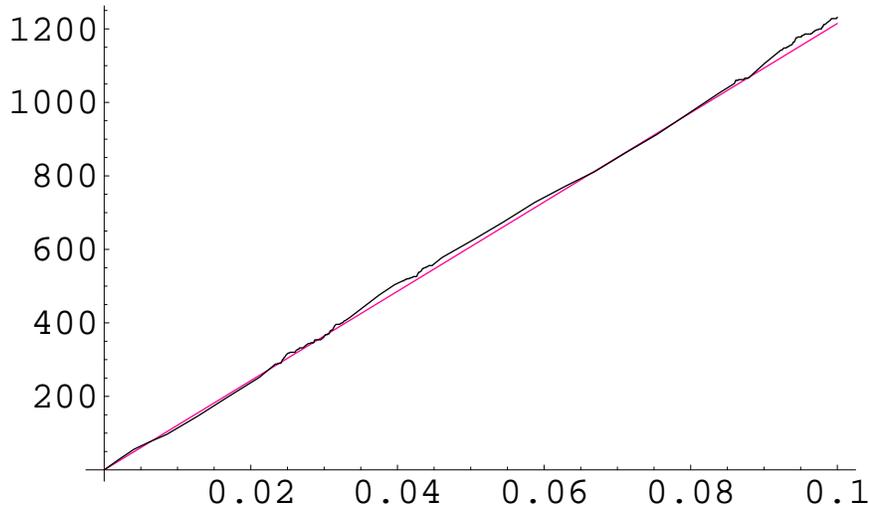,width=0.7\textwidth}
\end{center}
\caption{The number of vacua ${\cal N}(-F,F)$ plotted against $F$.
The actual distribution for our example with $N=15$ is the bumpy
line, compared with the statistical expectation}
\end{figure}

\begin{figure}
\begin{center}
\epsfig{file=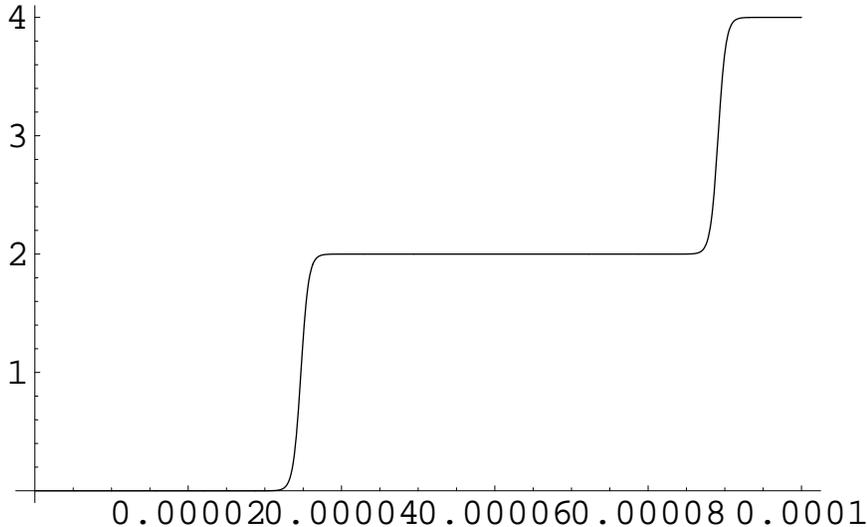,width=0.7\textwidth}
\end{center}
\caption{Closeup of the distribution--we see that the smallest $F$
is indeed of order $2^{-15}$.}
\end{figure}

\subsubsection{(Non)-scanning of couplings}
We are now in a position to ask whether the presence of the huge
$2^N$ number of vacua guarantees that we will be able to find one
with small enough vacuum energy. Interestingly, the answer is {\it
no}. To see why, let us examine the distribution for the vacuum
energy
\begin{equation}
\rho(\Lambda) = \frac{2^N}{\sqrt{2 \pi N} V_{{\rm dif}}}\,
e^{-\frac{(\Lambda - N \bar{V}_{{\rm av}})^2}{2 N V^2_{{\rm dif}}}}
\end{equation}
This distribution densely scans a region of width $\sim \sqrt{N}
V_{{\rm dif}}$ around a central value $\sim N V_{{\rm av}}$. This
would guarantee that we could find a tiny vacuum energy of order
$\sim M_*^4 2^{-N}$, for $\bar{V}_{\rm av}$ of order $\sim
1/\sqrt{N} M_*^4$. But what should we expect for $\bar{V}_{{\rm
av}}$? Unfortunately in a non-supersymmetric theory, the vacuum
energy is UV sensitive and not calculable. However, we have $N$
non-supersymmetric sectors, each of which contributes at least an
amount $\sim M_*^4$ to the vacuum energy, so it is reasonable to
expect that
\begin{equation}
N \bar{V}_{\rm av} = \bar{\Lambda} \sim N M_*^4
\end{equation}
and so $\bar{V}_{{\rm av}} \sim M_*^4$. One might naively think that
the signs of the vacuum energy randomly vary from sector to sector,
so that the sum of vacuum energies fluctuates around zero, and
$\bar{\Lambda}$ is of order $\sqrt{N}$ rather than $N$. However, in
a non-supersymmetric theory, there is nothing special about zero
vacuum energy, so there is no reason to expect the vacuum energy to
scan around zero.  For instance, the zero-point energy of the $N$
scalars is $\sim NM_*^4$, so that is where one should expect the
mean vacuum energy to lie; a much smaller value would require a
tuning.  If $\bar{\Lambda}$ has its natural value near $\sim N
M_*^4$, then in the central region of the Gaussian distribution,
values of the vacuum energy with $\Lambda = \bar{\Lambda} \pm \delta
\Lambda$ are finely scanned, with
\begin{equation}
\delta \Lambda/\bar{\Lambda} \sim \frac{1}{\sqrt{N}}
\end{equation}
In particular, this tells us that we {\it don't} get to scan around
zero vacuum energy in the central region of the Gaussian.

This small width for the scanned region for the vacuum energy has a
purely statistical origin; the actual width of the scanned region
can be even further suppressed by weak coupling factors. Let us
suppose that the quartic coupling for our scalars $\phi$ is perturbative
$\sim g^2$, and all the other couplings in the theory are included
with their natural sizes. Note that $V_{dif}$ vanishes in the limit
where the cubic and linear couplings vanish, so we include these
with their {\it largest} natural sizes; the potential $V(\phi)$ is
then of the form
\begin{equation}
V(\phi) = g^2 \left(\phi^4 + \epsilon M_* \phi^3 + \epsilon^2 M_*^2
\phi^2 + \epsilon^3 M_*^3 \phi \right)
\end{equation}
where $\epsilon^2 \sim 1/(16 \pi^2)$ is a loop factor. The vevs of
$\phi$ are then of order $\langle \phi \rangle \sim \epsilon M_*$,
so
\begin{equation}
V_{{\rm dif}} \sim g^2 \epsilon^4 M_*^4
\end{equation}
and the tunneling rate for the false vacuum can easily be seen to be
of order
\begin{equation}
\Gamma \sim M_*^4  e^{- c \frac{27 \pi^2}{g^2}}
\end{equation}
where $c$ is an ${\cal O}(1)$ factor. Thus we need to be
parametrically at weak coupling, with small $g^2/16 \pi^2$,  to have
a long-lived enough vacuum. In this model, $\delta \Lambda \sim
\sqrt{N} V_{{\rm dif}} \sim \sqrt{N} g^2 \epsilon^4 M_*^4$, while
$\bar{\Lambda} \sim N \epsilon^2 M_*^4$, and
\begin{equation}
\delta \Lambda/\bar{\Lambda} \sim \frac{g^2}{16 \pi^2}
\frac{1}{\sqrt{N}}
\end{equation}
so as promised, we see that, in addition to the statistical
$1/\sqrt{N}$ suppression of the scanning for the vacuum energy,
there is also a weak-coupling suppression by $\sim g^2/16 \pi^2$,
which is tied to the longevity of the false vacuum in this model.

While for natural values of $\bar{\Lambda}$, the large number $2^N$
of vacua does not in itself guarantee the presence of a vacuum with
tiny vacuum energy in the fat region of the Gaussian distribution,
it may be possible to find one on the tail of the Gaussian, since
\begin{equation}
\rho(\Lambda \sim 0) \sim \frac{1}{V_{{\rm dif}}} \times \left(2 \,
e^{-\frac{\bar{V}_{{\rm av}}^2}{V_{{\rm dif}}^2}} \right)^N
\end{equation}
Therefore if
\begin{equation}
\frac{\bar{V}_{{\rm av}}^2}{V_{{\rm dif}}^2} < {\rm log} ~2
\end{equation}
we can still find a tiny vacuum energy at large $N$. However, this
is condition may be difficult to satisfy; for instance, in our
weakly coupled coupled model,
\begin{equation}
\frac{\bar{V}_{av}^2}{V_{{\rm dif}}^2} \sim \frac{16 \pi^2}{g^2} \gg
{\rm log} 2
\end{equation}

We have come to an interesting conclusion. In order for our
landscape to solve the cosmological constant problem, in addition to
the exponentially large number of vacua, we require an extra
accident! Either a tiny CC can arise on the tail of a Gaussian
distribution, or more likely, an extra accidental fine-tuning is
required for the central value of $\bar{\Lambda}$ to be of order
$\sim \sqrt{N} M_*^4$ rather than $\sim N M_*^4$.

However, if this accident did not happen, our landscape would not
lead to a meaningful low-energy effective theory. One might imagine
that, in a larger ``global" picture of the landscape, in most
regions $\Lambda$ never gets scanned enough to find a small enough
expansion rate for the universe to allow the formation of structure,
but that in $\sim 1/\sqrt{N} \sim$ a few percent of the regions, the
required accident occurs and leads to a low-energy effective theory
with a finely scanned $\Lambda$. Also, as we will see, in
supersymmetric theories the situation is better--the combination of
SUSY together with unbroken (discrete) R symmetries can guarantee
that the distributions are peaked around $\Lambda = 0$.

\subsubsection{Coupling to the Standard Model}
We can now imagine coupling the landscape fields to the Standard
Model. At the renormalizable level, the only allowed couplings are of
the form
\begin{equation}
\mu \phi h^\dagger h, \lambda \phi^2 h^\dagger h
\end{equation}
which would suggest that, together with the vacuum energy, the Higgs
mass is scanned from vacuum to vacuum (actually, integrating out the
heavy $\phi$ excitations around the vacua at tree-level also
generates a Higgs quartic that varies).  This would be the case if
the vevs $\langle \phi \rangle$ are much smaller than the cutoff
$M_*$ of the theory, but in our non-supersymmetric examples this is
not natural. We can't neglect higher-dimension operators, which are
an expansion in powers of $(\phi/M_*)$, so that the Lagrangian takes
the form
\begin{eqnarray}
{\cal L} &=&  - \left(g_0^{-2} +  \sum_i
g_i^{-2}(\frac{\phi_i}{M_*})\right) F_{\mu \nu}^2 + \left(Z_{0 \psi}
+ \sum_i Z_{\psi i}(\frac{\phi_i}{M_*}) \right) \bar{\psi}
\bar{\sigma} D \psi \nonumber \\ &+& \left(Z_{h 0} + \sum_i Z_{h
i}(\frac{\phi}{M_*}) \right) |D h|^2 - \left(m_0^2 +
m^2_i(\frac{\phi_i}{M_*}) \right) |h|^2 -
\left(\lambda_0 + \sum_i \lambda_i(\frac{\phi_i}{M_*}) |h|^4 \right) \nonumber \\
&+& \left(\lambda_{0 ab} + \sum_i \lambda_{ab
i}(\frac{\phi_i}{M_*})\right) \psi_a^c \psi_b h + \cdots
\end{eqnarray}
The notation is self-explanatory. For the case of the Yukawa
couplings, we will assume for now that the $\phi$'s are not
charged under any of the approximate chiral symmetries of the SM
fermions, so that whatever explains the Yukawa hierarchies and
suppresses the $\lambda_{0 a b}$ (via flavor symmetries,
separation in extra dimensions, or whatever else) also similarly
suppresses $\lambda_{ab i}$.

Note that we have again assumed that there are no cross-couplings in
the interactions of the SM with the landscape fields. As in the case
with the vacuum energy in the last subsection, some such couplings
are inevitably induced by loops of gravity, and now also loops of SM
fields. But just as before, the cross-couplings are parametrically
suppressed at large $N$ and can be neglected.

In this non-supersymmetric example, the vevs $\langle \phi_i
\rangle$ will be close to $M_*$, and the suppression of higher
dimension operators is not significant, so all the couplings can in
principle scan in the landscape. Actually, if the theory is weakly
coupled with a weak coupling factor $\sim g^2$, the expansion in
higher dimension operators is really in the combination $(g
\phi)/M_*$, so that the higher dimension operators will indeed be
suppressed. However, as we have already seen for the vacuum energy,
even discounting any weak coupling factors, the variation in any
coupling $c$ is generally small for purely statistical reasons
\begin{equation}
\delta c/\bar{c} \lsim \frac{1}{\sqrt{N}}
\end{equation}
The reason is just as for the vacuum energy. The value of a coupling
$c$ for any given vacuum $\{\eta\}$ is of the form
\begin{equation}
c_{\{\eta\}} = c_0 + \sum_i c_{{\rm av} i} + \sum_i \eta_i c_{{\rm
dif i}}
\end{equation}
which are significantly scanned in the landscape in a region
\begin{equation}
c_0 + N \bar{c}_{{\rm av}} \pm \sqrt{N} c_{{\rm dif}}
\end{equation}
Now, the magnitude of the variation of couplings depends upon how
the couplings $c_i(\phi_i/M_*)$ to the landscape fields are scaled
at large $N$. If the scaling is chosen so that the large $N$ limit
has fixed coupling strengths, then $N \bar{c}_{{\rm av}}$ is
subdominant to $c_0$, and
\begin{equation}
\delta c/\bar{c} \ll \frac{1}{\sqrt{N}}
\end{equation}
On the other hand, if the $N \bar{c}_{{\rm av}}$ term dominates
$c_0$, then we have the largest scanning region, which is still
suppressed at large $N$
\begin{equation}
\delta c/\bar{c} \sim \frac{1}{\sqrt{N}}
\end{equation}

\subsubsection{Origin of weak couplings $\alpha \sim 4 \pi/N$}

It is interesting to consider this last limit in more detail.
Suppose that the theory is intrinsically strongly coupled at the
scale $M_*$. From naive dimensional analysis \cite{NDA}, we expect
that the effective Lagrangian takes the form
\begin{equation}
{\cal L} = \frac{1}{16 \pi^2} \left( \sum_i f_i(\frac{\phi_i}{M_*})
F_{\mu \nu}^2 + \cdots \right)
\end{equation}
where all the $f_i$ are of ${\cal O}(1)$. Then, since each of the terms
in the sum is of ${\cal O}(1)$, the Lagrangian in any of the
densely scanned vacua takes the form
\begin{equation}
\sim \frac{N}{16 \pi^2} \left((1 \pm {\cal O}(1/\sqrt{N})) F_{\mu
\nu}^2 + \cdots \right)
\end{equation}
(Here again, the fact that there is nothing special about zero tells
us that the expected coefficient is of order $N$, not $\sqrt{N}$).
This means that all the dimensionless couplings in the theory are
weak
\begin{equation}
\frac{1}{g^2} \sim \frac{N}{16 \pi^2} \to \frac{g^2}{16 \pi^2} \sim
\frac{1}{N}
\end{equation}
and of course none of them scan significantly
\begin{equation}
\delta g/g \sim \frac{1}{\sqrt{N}}
\end{equation}
This can provide an attractive explanation for the relative weakness
of the SM couplings, with $\alpha \sim 4 \pi/N$.

\subsection{General lessons}
We have seen that, in our landscape of vacua, despite the presence
of a huge number $2^N$ of vacua, the most natural situation is that
{\it none} of the couplings are effectively scanned from vacuum to
vacuum. This is because while the scanning allows us to sample a
region of size proportional to $\sqrt{N}$ around a mean, the mean
coupling induced from interactions with $N$ landscape fields will
itself be proportional to $N$, so that the relative deviations from
central values are down by (at least) ${\cal O}(1/\sqrt{N})$.

\begin{figure}
\begin{center}
\epsfig{file=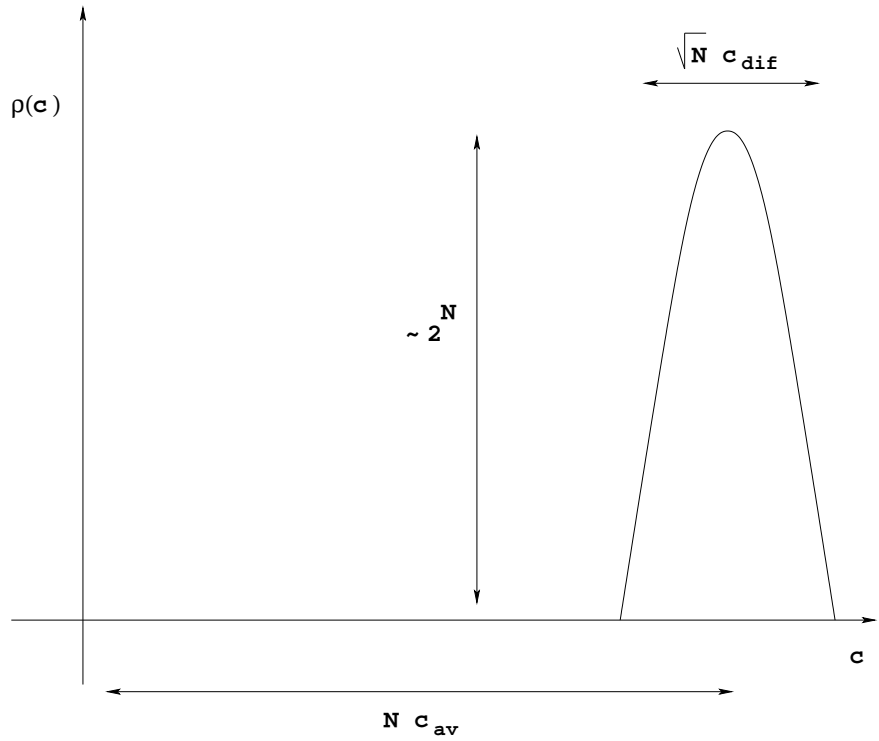,width=0.7\textwidth}
\end{center}
\caption{Generic scanning of couplings on our landscape. Note that
the region of the scanned couplings is small, $\delta c/c \lsim
1/\sqrt{N}$, since the mean value of the coupling grows as $N$ while
the width only grows as $\sqrt{N}$}
\end{figure}

This conclusion can be avoided if, for some symmetry reason, the
central value of the coupling $\bar{c}_{av}$ vanishes. In this case,
a region of size $\sqrt{N} c_{{\rm dif}}$ is efficiently scanned.
\begin{figure}
\begin{center}
\epsfig{file=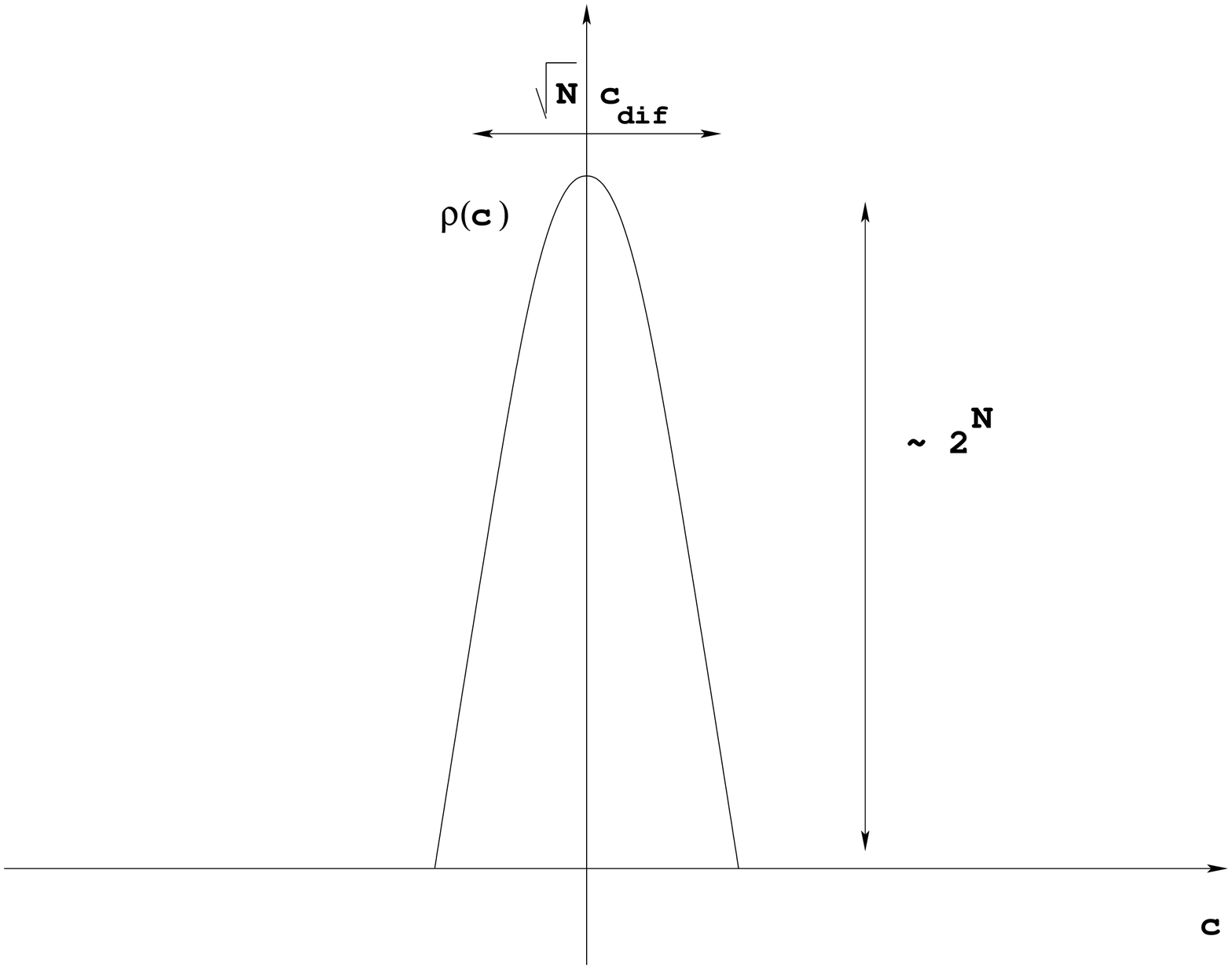,width=0.7\textwidth}
\end{center}
\caption{If for a symmetry reason, the mean value of the coupling
vanishes, the scanning can be efficient.}
\end{figure}
However, in our non-supersymmetric examples, there are no symmetries
that can guarantee $\bar{c}_{{\rm av}} = 0$. Even the $\phi \to -
\phi$ symmetry is necessarily broken in order to have $V_{{\rm dif}}
\neq 0$, required to be able to scan the vacuum energy.

This means that, naturally, none of the couplings vary significantly
in the landscape, unless for an accidental reason, the central
values are of order $\sqrt{N}$ rather than $N$. This would take an
accident or tuning of order $1/\sqrt{N}$.  But this accident {\it
must} happen for the relevant operators in theory, such as the
cosmological constant and the Higgs mass. If not, there would not be
any low-energy effective theory to speak of: if the CC doesn't scan,
all the vacua have curvatures near the fundamental scale, and if the
Higgs mass doesn't scan, there is no Higgs field in the low-energy
effective theory.  In this sense, the relevant operators are
special, since they have the largest impact on the IR physics. But
there is no worry about {\it all} parameters varying wildly on the
landscape--on the contrary they naturally don't vary significantly
at all.

We have therefore provided a rationale for the existence of a
``friendly neighborhood" of the landscape, where the only couplings
that are scanned are relevant operators, while the dimensionless
couplings are essentially fixed. If our low-energy model for the
landscape could be derived from a UV complete theory, despite the
enormous number of vacua in the effective theory, all the
dimensionless parameters could be predicted to ${\cal O}(1/\sqrt{N})
\sim$ 5 $\%$ accuracy each! And while the relevant couplings scan
and therefore take widely differing values across vacua, being
relevant, they have the largest impact on IR physics, and therefore
very gross environmental arguments (such as the existence of
structure or atoms) can be used to fix their values as well.

The theories where the large $N$ number of fields lead to weak
Standard Model couplings with $\alpha \sim 4 \pi/N$ are particularly
attractive, since in these cases, {\it all} hierarchies in nature can
be seen to be exponentially large in $N$. This is because the
asymptotically free couplings generate exponentially low scales
$\Lambda_{IR} \sim e^{-8 \pi^2/b g^2} \sim e^{-c N}$, and while the
relevant couplings are not determined, environmental arguments peg
their values to various powers of the scales $\Lambda_{IR}$. Thus
for instance, the ``atomic" principle fixes the weak scale (and thus
the Higgs mass squared parameter) to $\Lambda_{QCD}$, while
Weinberg's argument pegs $\Lambda^{1/4} \sim m_{DM}^2/M_{Pl}$, and
the $m_{DM}$ scale can be determined by dimensional transmutation.
Even in a theory without DM, Weinberg's argument would have set
$\Lambda^{1/4} \sim \eta_B \Lambda_{QCD}$. Thus, in this picture of
the world, all hierarchies are determined first through dynamics
(via dimensional transmutation of marginal couplings) and then by
environmental arguments (for the relevant couplings) to be of the
form $e^{-c N}$ for various $O(1)$ parameters $c$.

\subsection{Supersymmetric Landscapes}

We can also build supersymmetric landscapes. This gives an immediate
advantage--what were UV sensitive relevant operators in the SM, such
as the cosmological constant and the Higgs mass, are UV insensitive
in supersymmetric theories, so our distributions for the CC and
scalar masses will be calculable.

In analogy with our non-SUSY example, suppose we have a chiral
superfield with a general cubic superpotential
\begin{equation}
W = m^2 \phi + \mu \phi^2 + \lambda \phi^3
\end{equation}
where we take $m \sim \mu$ near the cutoff scale $M_*$. There are
two supersymmetric minima at $\langle \phi \rangle= \phi_{\pm}$, and
at these two minima, there are in general two different vevs for the
superpotential $\langle W \rangle = W_{\pm}$. Once again imagine we
have $N$ decoupled sectors of this type, so there are $2^N$
supersymmetric minima labeled by $\{\eta\}$. The value of the
superpotential in these minima is
\begin{equation}
W_{\{\eta\}} = \sum_j W_j = N \bar{W}_{{\rm av}} + \sum_i \eta_i
W_{{\rm dif} i}
\end{equation}
As in the previous section, we will now have a distribution of
possible $W$'s, peaked around $N W_{{\rm av}}$ with a width of order
$\sqrt{N} W_{{\rm dif}}$. When we turn on gravity, each of these
become AdS minima, with vacuum energy $-3 |W|^2/M_{Pl}^2$
distributed around $-3|N W_{{\rm av}}|^2/M_{Pl}^2$.

Suppose now that SUSY is broken in a hidden sector, at a scale
$\Lambda_S$, parametrically smaller than $M_*$, contributing a
positive vacuum energy $\sim \Lambda_S^4$. Again, for the most
natural value of $W_{{\rm av}} \sim M_*^3$, it is not possible to
cancel this positive energy from the $-3 |W|^2/M_{Pl}^2$ AdS vacuum
energies.

However, in this supersymmetric theory, there is a symmetry that can
enforce $\bar{W}_{{\rm av}} = 0$. Suppose that the superpotential is
an {\it odd} polynomial in $\phi$, e.g.
\begin{equation}
W = \lambda \phi^3 - \mu^2 \phi
\end{equation}
This form can be enforced by a discrete $Z_4$ R symmetry under which
$\phi(y, \theta) \to - \phi(y, i \theta)$. In this case, $W_+ = -
W_-$, so that $W_{{\rm av}} = 0$. Then, the landscape sector scans
vacuum energies in the range
\begin{equation}
0 \ge \Lambda_{{\rm landscape}} \gsim -3\frac{|\sqrt{N} W_{{\rm
dif}}|^2}{M_{{\rm Pl}}^2}
\end{equation}
so that it is always possible to find a vacuum that cancels the $+
\Lambda_S^4$ vacuum energy from SUSY breaking, to an accuracy of
order $\sim 2^{-N} M_*^4$.

Depending on how this landscape sector couples to the Standard Model
fields, we can arrive at a variety of low-energy theories.

\subsubsection{MSSM scanning only $\Lambda$}
The most straightforward possibility is that the landscape sector
does not couple to the MSSM, or that, as in our non-SUSY examples,
there is a coupling, but none of the MSSM couplings scan effectively
in the landscape. In this case, the only coupling that does scan
effectively is the cosmological constant. Thus, this is a landscape
that gives us the conventional supersymmetric standard model at
low-energies, with the landscape only playing a role in solving the
cosmological constant problem.

\subsubsection{The  $\mu$ and doublet-triplet splitting problem}
Another interesting possibility arises if we recall that there is
another natural parameter in the MSSM that can be set to zero by
symmetry arguments: the $\mu$ term $\mu H_u H_d$ in the
superpotential. In fact, we can use exactly the same $Z_4$ R
symmetry as before, with the $H_{u,d}$ carrying R charge $0$ and the
SM matter fields carrying $R$ charge 1, to forbid a $\mu$ term.
However, we can have couplings of the form
\begin{equation}
\sum_i g_i M_* P_i(\phi_i/M_*) H_u H_d
\end{equation}
where $P_i$ is any odd polynomial in $\phi_i$. In turn, this leads
to the $\mu$ term being scanned about a zero central value on the
landscape. Note that the discrete $R$ symmetry forces the Yukawa
couplings to be of the form
\begin{equation}
\sum_i \lambda_{a b i} (\phi_i/M_*) f_a f_b^c H
\end{equation}
where the $\lambda_{a b i}$ are {\it even} in $\phi$, so they are
not effectively scanned in the landscape.

Therefore in this landscape, it is the CC and $\mu$ term that are
scanned, while all other parameters (including dimensionless
couplings and SUSY breaking soft terms) are fixed. This can in turn
lead to an environmental solution to the $\mu$ problem, if we use
the atomic principle to fix the Higgs VEV, which is applicable
independent of the mechanism of SUSY breaking (gravity, gauge etc.
mediation), and does not require the presence of any light
particles. This is especially interesting if the landscape favors
very low energy scales of SUSY breaking, e.g. gauge mediated SUSY
breaking, as in these theories the $\mu$ problem is challenging and
typically requires  excessively clever model-building gymnastics and
a rich spectrum of new light particles. In grand unified theories
this mechanism evolves into a solution of the closely related
doublet-triplet splitting problem \cite{dg}.  In minimal SU(5), for
example, the superpotential
\begin{equation}
W = H_u\, \left( M + \langle \Sigma \rangle \right)\,H_d
\end{equation}
responsible for the Higgs doublet and (SU(3)) triplet masses (where
$\langle \Sigma \rangle$ is the adjoint VEV $\sim M_{GUT} \sim
10^{16}$ GeV ) typically gives a mass of order $\sim M_{GUT}$ to
both the doublets and triplets. With $M$ scanned, only where $M$ and
$\langle \Sigma \rangle$
 almost cancel---to give a small $\mu$ term for the doublets---
is the electroweak VEV near its experimental value, as required by the atomic principle.

\subsubsection{The Standard Model, scanning only $\Lambda$ and
$m_h^2$}

Suppose that the SUSY breaking masses in the MSSM, $\tilde{m} \sim
F_Z/M_*$, are much much higher than the TeV scale, but still small
compared with the scale $M_*$. What does the effective theory look
like at energies far lower than $\tilde{m}$? Of course, in most of
the landscape, all the scalars (including both Higgses) as well as
the gauginos will be near $\tilde{m}$. Since the $\mu$ term for the
Higgs is scanned, it is possible to find Higgsinos surviving beneath
$\tilde{m}$. However, there is a different part of the landscape,
where the squarks, sleptons and gauginos are heavy, but a single
linear combination of the Higgses can be light. Recall that the
Higgs mass matrix is of the form
\begin{equation}
\pmatrix{\tilde{m}_u^2 + \mu^2 & \tilde{m}^2_{ud} + B \mu \cr B \mu
+ \tilde{m}^2_{ud} & \tilde{m}_d^2 + \mu^2}
\end{equation}
As mentioned, $\tilde{m}^2_{u,d,ud}$ and $B$ do not vary in the
landscape, while $\mu$ is scanned.  Note that the sum of the
eigenvalues of the $m^2$ matrix is
\begin{equation}
m_1^2 + m_2^2 = \tilde{m}_{u}^2 + \tilde{m}_d^2 + 2 \mu^2 >
\tilde{m}_u^2 + \tilde{m}_d^2
\end{equation}
and if $\tilde{m}_u^2 + \tilde{m}_d^2 > 0$, then at least one of the
eigenvalues of $m^2$ must be comparable to $\tilde{m}^2$. However,
as $\mu$ varies in the landscape, it is possible that one of the
eigenvalues is fine-tuned to be light. Note that clearly this can
only happen in the region of the landscape where $\mu \sim
\tilde{m}$, and that in this region the Higgsinos are heavy, also at
$\tilde{m}$. At low-energies, we have the Standard Model with a
single Higgs doublet. The value of $\mu$ where the SM Higgs is light
can be found just by setting the determinant of the mass matrix to
zero. This fixes $\mu$ and also the linear combination of $h_{u,d}$
that is tuned to be light
\begin{equation}
h = \mbox{cos} \beta h_u + \mbox{sin} \beta h_d
\end{equation}
Note that as $\mu$ is then scanned in this vicinity, $m_h^2$ is
clearly scanned, but tan$\beta$ remains fixed to an accuracy of
$O(\sim m_h^2/\tilde{m}^2)$.

We have thus found an example of a SUSY landscape that reduces to
the Standard Model at energies far below a (very high) SUSY breaking
scale $\tilde{m}$, and where a discrete symmetry guarantees that the
vacuum energy and Higgs mass are finely scanned, while all other
parameters are effectively fixed.

\subsubsection{Scanning SUSY breaking scales}

In our discussion so far, we imagined that the SUSY breaking scale
was not scanned in the landscape. It is also instructive to consider
theories where the SUSY breaking sector couples to the landscape.
Consider a chiral superfield $X$, with a coupling to the landscape
sector of the form
\begin{equation}
W_X = \sum_i M_*^2 \lambda_i P_i(\phi_i/M_*) X
\end{equation}
where again the $Z_4$ R symmetry guarantees $P_i$ is an odd
polynomial in $\phi_i$. Integrating out the heavy $\phi$'s would
generate a superpotential for $X$ of the schematic form
\begin{equation}
M_*^2 (\lambda X) + M_* (\lambda X)^2 + (\lambda X)^3 + \cdots
\end{equation}
which clearly has stationary points for
\begin{equation}
X \sim \frac{M_*}{\lambda}
\end{equation}
However, higher order terms in the Kahler potential for $X$
\begin{equation}
\int d^4 \theta - \frac{X^2 X^\dagger}{M_*} + \mbox{h.c.} + \cdots
\end{equation}
can lead to a local minimum with broken SUSY, with SUSY broken by
non-vanishing $F_X$. Because of the $Z_{4}$ R symmetry, $F_X$ will
vary around a mean of zero,
\begin{equation}
\frac{F_X}{M_*^2} = \sum_i \eta_i P_{{\rm dif} \, i}
\end{equation}
and will scan effectively in the landscape. Since the distributions
of $F_X$ and $W$ are independent, the tuning required to cancel the
CC is independent of the actual value of the SUSY breaking scale.

If we concentrate on the vacua with tiny vacuum energy, the physical
measure of SUSY breaking is the gravitino mass
\begin{equation}
m_{3/2}^2 = \frac{|F_X|^2}{M_{Pl}^2} = \frac{({\rm Re}F_X)^2 + ({\rm
Im} F_X)^2}{M_{Pl}^2}
\end{equation}
Both Re$(F_X)$ and Im $(F_X)$ have approximately flat distributions
around $0$. In turn, this implies that the density of distribution
for $m_{3/2}$ is proportional to $m_{3/2}$
\begin{equation}
\rho(m_{3/2}) \propto m_{3/2}
\end{equation}
That is, there are more vacua with ${\it high}$ scale SUSY breaking
in this theory. Note that if for some reason, all the $F_X$ in
the distribution had the
same phase (say because of a conserved $CP$ symmetry), we would
instead have $\rho(m_{3/2}) \propto 1$. We can also generalize to
having $n$ fields $X_a$, $a = 1, \cdots, n$ that break SUSY. Then,
each of the Re($F_{X_a}$), Im($F_{X_a}$) have a flat distribution,
but
\begin{equation}
m_{3/2}^2 = \frac{\sum_a |F_{X_a}|^2}{M_{Pl}^2}
\end{equation}
has a distribution
\begin{equation}
\rho(m_{3/2}) \propto (m_{3/2})^{2n - 1}
\end{equation}%

Now, suppose the scalars of the MSSM also pick up a mass of order
$\tilde{m}^2 \sim m_{3/2}^2$ from gravity-mediated SUSY breaking. In
what fraction of vacua do we expect to have a Higgs mass of $\sim
m_h^2$? For a given $\tilde{m}^2 > m_h^2$, there is a tuning of
magnitude $\sim (m_h^2 / \tilde{m}^2)$ required to keep the Higgs
light. However, for $n > 1$, there are many more vacua with large
SUSY breaking scales, and the number of vacua with SUSY broken at
the scale $\sim \tilde{m}$ is proportional to
\begin{equation}
\label{realtune} \left(\frac{m_h^2}{\tilde{m}^2}\right) \times
\left(\frac{\tilde{m}^2}{M_*^2}\right)^{2n - 1}
\end{equation}
Thus, there are more vacua with a given value of the weak scale,
with SUSY broken at high scales than at low scales. The fact that
the distribution of $F$ and $D$ terms, if sufficiently uniform, will
{\it statistically} favor high scale SUSY breaking on the landscape
has been discussed in the context of stringy constructions in
\cite{lennyhigh,Douglashigh,DGT,Counter}.
It is even plausible that KK scale breaking
will be favored in this sense \cite{SaltSi}.

\subsubsection{A Landscape for split SUSY}

We can also build a landscape realizing the scenario of split SUSY.
In one limit of split SUSY, the scalars are only at about $\sim 100
- 1000$ TeV. This can arise from conventional hidden sector SUSY
breaking if there are no singlets in the hidden sector. While the
scalars can pick up a mass from operators suppressed by $M_* \sim
M_{GUT}$, so that $\tilde{m} \sim F/M_*$, the gauginos pick up a
mass $\sim \alpha/\pi m_{3/2}$ from anomaly mediation \cite{anom},
where $m_{3/2} \sim F/M_{Pl}$. So, we can have $\tilde{m} \sim 1000$
TeV, $m_{3/2} \sim 30$ TeV and the remaining fermions near the
$\sim$ TeV scale. This scenario is very plausible. We can realize it
on the landscape in the same way as our model above, which reduced
to the SM with a tuned Higgs mass at low energies.

But we can also consider the more extreme version of split SUSY,
with the scalars and gravitino being very heavy. To do this, we want
to break SUSY without breaking R symmetry.

Fortunately, the classic model of SUSY breaking--the O'Raigheartaigh
theory, is precisely an example of a theory that breaks SUSY while
preserving $R$. Consider the theory with three chiral superfields
$X,Y,Z$ and superpotential
\begin{equation}
\lambda Y (Z^2 - \mu^2) + m Z X
\end{equation}
This model has an $R$ symmetry under which $Y,X$ have charge 2 while
$Z$ has charge 0.  It is simple to understand the physics in the
limit where $m < \lambda \mu$. The $Y$ equation of motion forces $Z$
to get a vev; there are two vacua with $Z = \pm \mu$. In these
vacua, $Y,Z$ marry up to get a mass $\mu$. Integrating out the heavy
fields leaves us with the superpotential
\begin{equation}
W = m \mu X
\end{equation}
while at 1-loop, we generate a correction to the Kahler potential of
$X$ as
\begin{equation}
\Delta K = -\frac{1}{24 \pi^2} \left|\frac{\lambda m}{\mu} \right|^4
\frac{(X^\dagger X)^2}{|\mu|^2}
\end{equation}
This breaks SUSY with $F_X = m \mu \neq 0$, while the induced term
in the Kahler potential stabilizes the modulus $X$ around $X = 0$
with the correct sign so that $m_X^2 > 0$. Thus, SUSY is broken
while $R$ is preserved.

It is a simple matter to extend this picture to a landscape. We take
$N$ copies of $Y,Z$ as
\begin{equation}
W = \lambda_j Y_j (Z_j^2 - \mu_j^2) + m_j Z_j X
\end{equation}
This form of the superpotential can be guaranteed by $R$ symmetries
and a discrete $Z_2$ symmetry under which $Z_j$ and $X$ change sign.

Once again, there are $2^N$ vacua. In each of these vacua, SUSY is
broken and $R$ symmetry is preserved, and
\begin{equation}
F_X = m_j \langle Z_j \rangle
\end{equation}
Note that $F_X$ is scanned around a mean value $F_{X {\rm av}} = 0$
with a width of order $\sim \sqrt{N} M_*^2$; this is guaranteed by
the $Z_2$ symmetry. In this model, it is therefore {\it positive}
vacuum energies $\Lambda = |F_X|^2$ that are scanned in the
landscape over a range
\begin{equation}
N M_*^4 \gsim \Lambda \ge 0
\end{equation}

Now, in order to find a vacuum with small cosmological constant, we
need to have a non-zero vev of the superpotential, which requires a
source of $R$-breaking. Suppose that there is a sector with pure
SUSY Yang-Mills with dynamical scale $\Lambda_{{\rm strong}}$,
giving rise to gaugino condensation with
\begin{equation}
\langle W \rangle = \Lambda_{{\rm strong}}^3
\end{equation}
Recall that the gauge couplings and hence $\Lambda_{{\rm strong}}$
do not scan significantly in the landscape. If we have many groups
with gaugino condensation, we will get the sum of $\Lambda_{{\rm
strong}}^3$ from each sector, and $W$ will be dominated by the
largest $\Lambda_{{\rm max}}$.

Therefore, while the distribution for $F_X$ does not favor either
low or high energy SUSY breaking, the requirement of cancelling the
cosmological constant fixes the SUSY breaking scale to be
\begin{equation}
|F_X| \sim \frac{\Lambda_{{\rm max}}^3}{M_{Pl}}
\end{equation}
We stress that this is not simply a purely statistical question
about there being more or less vacua with a given scale of SUSY
breaking and vanishing vacuum energy--it is {\it only} possible to
cancel the vacuum energy for a {\it fixed} value of SUSY breaking
scale. If $\Lambda_{{\rm max}}$ is relatively high, this will {\it
force} (rather than merely statistically favor) high-energy SUSY
breaking. This landscape illustrates one of the motivations for
split SUSY given in \cite{AD}. There may be regions of the landscape
with SUSY broken at low energies, and others with SUSY broken at
high energies. The latter may naively be disfavored because of the
additional tuning required for the hierarchy problem. But, it may be
that it is {\it impossible} to cancel the cosmological constant in
theories with a conventional low breaking scale for SUSY. We see
this explicitly in our example.

Since the high-scale breaking of SUSY preserves $R$-symmetry, we are
naturally let to a prediction of split SUSY at low energies
\cite{AD,Splittwo}. Of course, we must also be able to scan for the
Higgs mass. Interestingly, in our landscape, of all the scalars of
the MSSM, {\it only} a single Higgs doublet can be fine-tuned to be
light. Note that the $R$-symmetry forbids any $\mu$ term for the
Higgsinos; however, a $B$-term of the form
\begin{equation}
\int d^2 \theta b_j \frac{Z_j X}{M_*}  H_u H_d
\end{equation}
is allowed, and scans around $0$ in the landscape. Meanwhile, the
soft masses for all the scalars are essentially fixed. This means
that none of the squarks or sleptons can become light. But the Higgs
mass matrix is
\begin{equation}
\pmatrix{\tilde{m}_u^2 & \mu B \cr \mu B & \tilde{m}_d^2}
\end{equation}
which can have a single small eigenvalue over the appropriate range
of $\mu B$'s.

Thus, we have SUSY broken at a very high energy scale, with all
scalars heavy and a single finely tuned Higgs light, and light
Higgsinos and gauginos. Of course, the inevitable breaking of R by
$\langle W \rangle$ eventually generates $R$-breaking fermion mass
terms, leading to light but not massless Higgsinos and gauginos. As
discussed in \cite{AD,Splittwo}, this can in fact naturally lead to
gauginos and Higgsinos near the $\sim 100$ GeV scale.

\subsection{Relation to the string theory landscape}

The motivation for the idea that we live in a ``friendly
neighborhood" of the landscape, where only the dimensionful
(relevant) couplings of the theory are finely scanned from vacuum to
vacuum, comes entirely from data via Weinberg's prediction of the
cosmological constant. Our simple field theory examples above
illustrate what such a landscape might look like. These models are
meant to only describe a ``local" region of the landscape where
things like spacetime dimensionality, low-energy gauge symmetry,
particle content and so on are fixed. There is of course a much
larger ``global" landscape, where all of these things can change. It
is important to stress that are also other ``local" landscapes where
``friendliness" is not guaranteed in the same way. Let us call the
landscapes we have studied $2^N$ landscapes for obvious reasons.
Some of the conclusions which follow from basic statistics applied
to the $2^N$ ensemble, do not apply to slightly more complicated
(but still well-motivated) ensembles. The best studied case in
string theory so far, is the ensemble of type IIB Calabi-Yau flux
vacua, which exhibits some differences.

As we have emphasized, one of the main features of the $2^N$
landscapes  is that one finds that scanning of couplings is {\it
difficult}. In particular, we find Gaussian distributions which make
it difficult to scan even small vacuum energy, under some
assumptions; or in SUSY theories, a Gaussian distribution of $W$
values.

In the IIB flux ensemble, the distributions of $W$ values and vacuum
energies (under suitable assumptions, but really governed by periods
of the Calabi-Yau compactifications) have also been studied in some
detail now \cite{DD,DDtwo}. The results are somewhat different --
for instance, $W$ does not have a Gaussian distribution, and
scanning a large range of values is not difficult, even without the
microscopic imposition of an R-symmetry.  This is certainly not a
contradiction since the ensemble involves more complicated
interactions between the fields -- the periods are generally
complicated transcendental functions of the moduli, and hence a
distribution governed by randomly selected flux potentials need not
behave like the $2^N$ ensemble.

However,  this illustrates that even very basic results for vacuum
distributions may vary in different plausible ensembles. In
non-supersymmetric $2^N$
ensembles, nothing is efficiently scanned.  The minimal assumption
that by luck $\Lambda$ can be scanned then leaves one in a friendly
neighborhood.  Or in SUSY examples with a discrete R symmetry on the
landscape, the scanning of $\Lambda$ (and the Higgs mass) can be
guaranteed by the discrete symmetry. In the IIB flux ensemble or
other ensembles that are being investigated in various corners of
string theory, the assumptions leading to friendliness may differ.
It may be even easier to scan $\Lambda$ while other quantities don't
vary much; or maybe in some unusual cases it will be hard not to
scan everything.

On the other hand, it is likely that ensembles
similar to the $2^N$ ensembles are actually
realized as part of the stringy landscape.  For instance,
one could presumably
design flux potentials with properties similar to our $2^N$ models by choosing
Calabi-Yau geometries with various discrete symmetries and local
singularities, where the symmetries and locality forbid strong
cross-coupling between the moduli related to different
singularities. A wide class
of effective field theories related to singularities in non-compact
Calabi-Yau geometries, realizing rather general Landau-Ginzburg effective
theories with many vacua (where the different vacua correspond to
flux vacua realized with different flux choices), can be found in \cite{GVW}.
It would be interesting to
demonstrate that brane constructions of the SM in these kinds of
Calabi-Yau models \cite{branecon}, can
be realized in a friendly neighborhood.  One
expects that varying fluxes far from the SM branes will scan over
such a friendly neighborhood in many cases.

Summarizing, our main point is {\it not} that all regions of the
landscape are indeed ``friendly". Rather, it is that {\it data}
seems to point to such a neighborhood, and that it is simple to
construct landscapes leading to friendliness. The non-trivial upshot
of all of this is to suggest a new, predictive framework for
thinking about physics beyond the standard model, which we turn to
next.

\section{Model-building rules in a predictive neighborhood}

We have argued that we live in a friendly neighborhood of an
enormous landscape of vacua, where only the relevant operators are
finely scanned from vacuum to vacuum, while the dimensionless
(nearly marginal) couplings are essentially fixed. This offers a
highly restrictive and predictive new framework for model-building,
where the finely tuned large hierarchies in nature are understood
from environmental considerations associated with gross large-scale
properties of the universe. These sorts of theories provide the
first known framework in which to simultaneously address the
cosmological constant problem together with the big and little
hierarchy problems.

Indeed the most minimal model along these lines is the Standard
Model itself, where the structure and atomic principles are used to
explain the fine-tunings for the cosmological constant and the weak
scale. As we have discussed above, note that one of the attractive
aspects of our predictive neighborhood of the landscape is that,
while there is no dynamical mechanism ``stabilizing" large
hierarchies, nonetheless the dimensionful quantities are
environmentally predicted to be exponentially small compared to the
cutoff, since they are inevitably pegged to other infrared scales
generated by dimensional transmutation. For instance, the atomic
principle sets the value of the weak scale to be close to
$\Lambda_{\rm QCD}$, and the exponential smallness of $\Lambda_{\rm
QCD}$ relative to the cutoff is understood via dimensional
transmutation.

But there are reasons to suspect that there is new physics beyond
the Standard Model near the TeV scale. Both the supersymmetric
unification of gauge couplings and the success of stable weakly
interacting stable TeV scale particles as dark matter candidates
suggests that the Standard Model must be augmented at accessible
energies. In the usual naturalness motivated theories of weak scale
SUSY, these successes are gained, while the atomic principle
explanation for the proximity of the weak to QCD scales is lost.
Furthermore, as we have discussed, the ``natural" theories all
suffer from a little hierarchy problem--why haven't we seen any
indirect evidence for new physics? This has been a growing tension
in beyond the standard model physics for the past two decades.

Split Supersymmetry offers a resolution to all of these puzzles.
SUSY is broken at very high energies, with all the scalars becoming
ultraheavy, except for single finely tuned Higgs that is forced to
be light due to the atomic principle. Yet, the fermions of the MSSM
are light, protected by an $R$ symmetry, and if they are near the
TeV scale, account both for Dark Matter as well as the successful
unification of gauge couplings. Since the only new particles beyond
the Standard Model are fermions with only Yukawa couplings to the
Higgs, there is an excellent reason why we have not seen any
indirect effects associated with these particles. All of the
dangerous operators in the MSSM are generated by light scalars in
1-loop diagrams, and disappear when the scalars are heavier than
$\sim 1000$ TeV. The only remaining indirect processes arise at two
loops with internal Higgs and new fermion lines; the most
interesting of these is an electric dipole moment for fermions which
is naturally just at the edge of planned new experiments
\cite{Splittwo}.

Split SUSY is attractive for its conceptual simplicity, as well as a
number of striking quantitative and qualitative predictions, which
include a long-lived gluino, and supersymmetric unification at a
single scale, of all 5 dimensionless parameters of the theory,
leading to 4 predictions analogous to gauge coupling unification
\cite{Splittwo}. One drawback is that, there is no a priori reason
to expect the gauginos and Higgsinos to be near the weak scale,
although remarkably, in some models for SUSY breaking the $\sim 100$
GeV scale arose rather naturally for their masses. But there is no
direct connection between the Dark Matter mass and the weak scale;
the apparent proximity of these would then be an accident.

\subsection{Minimal model for Dark Matter and Unification}

It is interesting to ask whether there are even more minimal
particle contents that give both a dark matter candidate and gauge
coupling unification. In examining split SUSY itself, it is easy to
see that at 1-loop the Higgsinos are most important for the relative
running of the couplings. Let us therefore consider a model with
only ``Higgsinos" at the TeV scale (where now this is just a
mnemonic for their quantum numbers since there is no longer a trace
of SUSY at low energies). The neutral component of the Higgsinos can
serve as a good dark matter candidate. We can also have additional
singlet fermions in the theory, with Yukawa couplings to the
``Higgsino"s and the Higgs. The DM particle can then be an admixture
of the singlet and neutral Higgsino components.

Remarkably, in this minimal model, the standard model gauge
couplings unify to high accuracy near $\sim 10^{14}$ GeV. While the
1-loop prediction for $\alpha_s(M_Z)$ is lower than in the MSSM,
two-loop running effects push up the predicted value for $\alpha_s$
to be about as far lower from central value as the MSSM prediction
is higher. The two-loop running couplings $\alpha_i^{-1}$ are
plotted below, with the Higgsinos put in at the mass of the top, and
the 1-loop weak scale threshold corrections for the top and SM gauge
bosons included in the usual way. We have {\it not} included the
two-loop running contributions from the Higgsinos themselves, as
these depend also on the Yukawa couplings to new singlet fermions.
Nonetheless, we expect that a more complete analysis will confirm
the basic result that gauge coupling unification works about as well
as in the MSSM.

\begin{figure}
\begin{center}
\epsfig{file=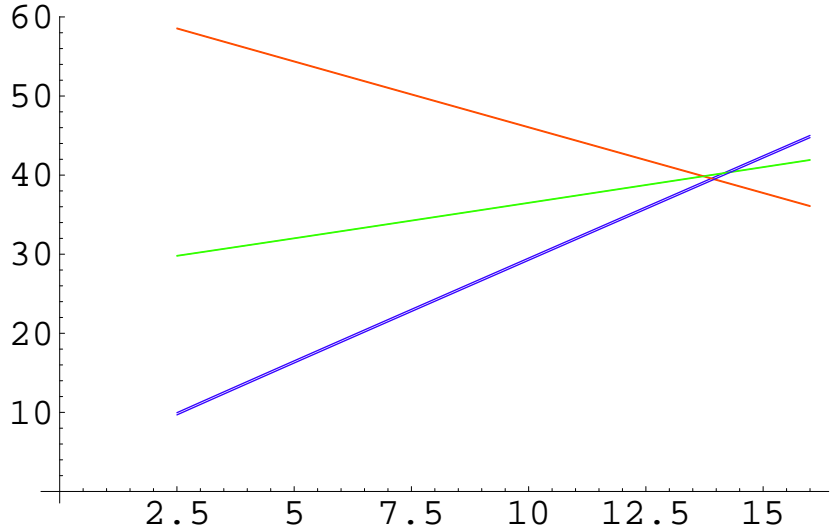,width=0.7\textwidth}
\end{center}
\caption{Two-loop gauge running of the couplings $\alpha_i^{-1}$ in
the minimal model, as a function of log$_{10}(E/$ GeV).}
\end{figure}

\begin{figure}
\begin{center}
\epsfig{file=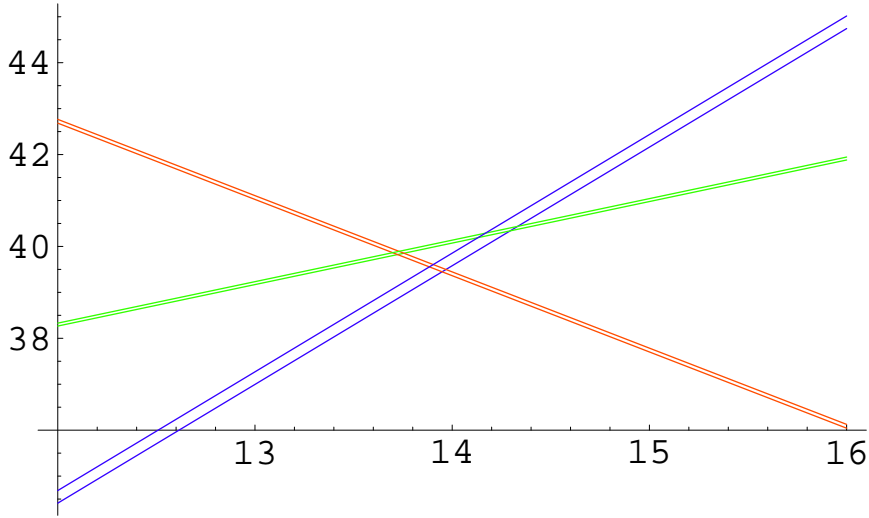,width=0.7\textwidth}
\end{center}
\caption{A close up of the running couplings near the unification
scale}.
\end{figure}

Since the unification scale is low,  we can't embed this theory into
standard grand unified models like $SU(5)$ or $SO(10)$, because of
gauge-mediated proton decay. However standard GUT's suffer from
problems of their own, such as doublet-triplet splitting, which are
most naturally resolved in higher-dimensional \cite{HN}(or
deconstructed \cite{Neal}) models with a GUT in the ``bulk" and
reduced gauge symmetry on the boundary. We can use exactly the same
ideas in our case to dispense with proton decay from $X,Y$ gauge
boson exchange. Indeed, having some of the extra dimensions in
string theory stabilized to moderately large sizes $\sim 10 - 100$
times larger than the string scale can lower the string scale close
to $\sim 10^{14}$ GeV, so it is natural to use these extra
dimensions for GUT breaking. Furthermore, these theories induce
calculable threshold corrections for the gauge couplings from
running between the compactification scale and the unification
scale, which are logarithmically enhanced relative the unknown UV
threshold corrections. In the case of split SUSY, this has recently
been studied \cite{philip}, and it has been found that models with
the Higgs doublets localized on a boundary further increase the
low-energy prediction for $\alpha_s$; it would be interesting to
redo this analysis for our model. If the third generation for the SM
is placed in the bulk (or on a boundary with an unbroken GUT
symmetry), we may preserve the success of $b/\tau$ unification
\cite{btau}, and also induce proton decay via CKM mixing--a simple
estimate suggests that this gives a rate for proton decay on the
edge of detection. A fundamental scale near $\sim 10^{14}$ GeV still
allows us to use higher dimension operators for neutrino masses, and
it is closer to the cosmologically allowed value of the axion decay
constant $f \sim 10^{12}$ GeV so that a fundamental-scale axion can
more easily solve the strong CP problem without having its coherent
oscillations overdominate the universe.

\subsection{New mechanisms for exponentially large hierarchies}

Weinberg's argument for the CC has a very satisfying {\it binary}
character to it--either the CC is bigger than a critical value, and
the universe is empty, or it is smaller, and the universe has
structure. The atomic principle explanation of the hierarchy does
not quite have the same character. However, our model-building rules
suggest new solutions to the hierarchy problem using even more
minimal environmental requirements than the existence of atoms,
which share the ``binary" feature of Weinberg's argument: unless the
Higgs mass is finely tuned to exponentially tiny values, an infrared
disaster happens, such as a universe devoid of any structure or
baryons.

We present below two models along these lines. Unlike split SUSY and
the ``only Higgsinos" model just discussed, these theories further
{\it predict} that Dark Matter is at the weak scale. Furthermore,
the minimal version of the second model we propose has the same SM
charged particle content as the ``only Higgsinos" model and
therefore has gauge coupling unification to high accuracy near $\sim
10^{14}$ GeV.

\subsubsection{Hierarchy from Structure}

Consider a toy universe with a massive $U(1)$ gauge field, and a
massive fermion, which consists of two Weyl spinors $\psi,\psi^c$ of
charge $+1$ and $0$, so that the mass term breaks the $U(1)$; there
are also other charged Weyl Fermions to cancel anomalies. The
massive fermion is stable, and the structure in this universe is
made from gravitational clumps made from this fermion.

The standard model of particle physics in this universe, would
include an elementary Higgs field $\phi$ with a mexican hat
potential, and a Lagrangian including
\begin{equation}
{\cal L} = {\cal L}_{kin} + y_t \phi \psi \psi^c - \frac{1}{4}
\lambda (\phi^\dagger \phi - v^2)^2
\end{equation}
Suppose $v$ is measured and found to be $\sim 10^{-17} M_{Pl}$.
Using the argument of \S1.1, we then immediately expect to measure a
cosmological term of magnitude $\Lambda^{1/4} \sim {v^2\over
M_{Pl}}$, but the smallness of $v$ relative to the cutoff would
remain mysterious. Theorists in this world might be tempted to use
e.g. low-energy supersymmetry to solve this hierarchy problem.

However, it is possible that the structure principle explains the
tininess of $v$ as well, so that both the cosmological constant and
hierarchy problems have a common solution.
As we have argued, in a predictive neighborhood of the landscape,
only the relevant operators (such as the cosmological constant and
the Higgs mass) are finely scanned, while dimensionless couplings
(nearly marginal operators) are essentially fixed. Thus, in
particular the Higgs quartic coupling does not change appreciably in
the landscape. We will now imagine that the quartic coupling at the
cutoff, $\lambda(M_*)$, starts {\it negative}
\begin{equation}
\lambda(M_*) < 0
\end{equation}
The quartic coupling is eventually be pushed positive under the
renormalization group, from the contribution to the RGE for
$\lambda$ from the Yukawa coupling $y_t$:
\begin{equation}
\frac{d}{dt} \lambda = - \frac{y_t^4}{16 \pi^2} + \cdots
\end{equation}
The quartic coupling will eventually cross zero and then run
positive at a scale $\Lambda_{\rm cross}$, which is exponentially
smaller than the cutoff $M_*$.

It is important to estimate the tunneling rate to the true vacuum,
given the negative quartic coupling. The situation is well
approximated by ignoring the mass term and just considering a theory
with a negative quartic potential $V(\phi) \sim -\lambda \phi^4$.
Naively there is no barrier, but in fact this really is a tunneling
problem and as we review in the appendix. The rate is exponentially
suppressed as $M^4 e^{-1/|\lambda|(M)}$, so that the vacuum is
long-lived enough as long as $|\lambda(M)|$ does not get too large
(see e.g. \cite{strumiavac} for a disucssion this in the context of
vacuum stability bounds on the Higgs mass in the Standard Model).

Now, let us see what the universe looks like for varying values of
$m^2$. First, suppose $m^2 > 0$. Then, there is no U(1) symmetry
breaking, and the fermions $t,t^c$ and the $U(1)$ gauge field are
massless. The field $\phi$ is of course massive, but is unstable to
decaying into the fermion. Thus, a cosmology in this universe will
be forever radiation dominated, with no hope of structure formation,
regardless of the value of $\Lambda$. Therefore, we can discard all
these possibilities using the structure principle.

For $m^2 < 0$, there are two possibilites. If $|m^2| > \Lambda_{\rm
cross}^2$, then since the quartic coupling is negative at the
appropriate scale $m$, there are simply no stable minima for the
potential, and the scalar rolls off to the cutoff scale. Again,
these can be discarded by the structure principle. Only for $|m^2| <
\Lambda_{\rm cross}^2$ does a stable minimum form.

We can see this explicitly by looking at the form of the quartic
potential around the point $\Lambda_{\rm cross}$ where the quartic
coupling vanishes; in this neighborhood we can approximate
$\lambda(\mu) = C$ log $\mu$ where $C$ is determined from the 1-loop
beta function, and therefore we can accurately approximate
\begin{equation}
V(\phi) = - C \phi^4 \mbox{log}\frac{\phi}{\mu_*} + m^2 \phi^2
\end{equation}
A graph of this function for various negative $m^2$'s shows that,
for $|m^2/C \Lambda_{\rm cross}^2| < .23$, we find a stable minimum
near $\phi \sim \Lambda_{\rm cross}$, while for larger $|m^2/C
\Lambda_{\rm cross}^2|$ there is no secondary minimum at all. The
position of the secondary minimum is always less than about $\langle
\phi \rangle_{max} \sim .2 \Lambda_{\rm cross}$.

\begin{figure}
\begin{center}
\epsfig{file=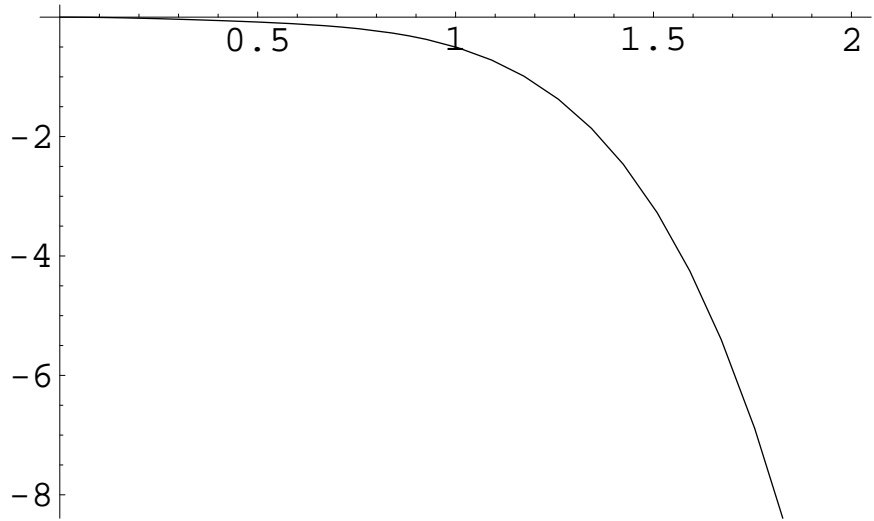,width=0.7\textwidth}
\end{center}
\caption{The effective potential as a function of $\phi/\Lambda_{\rm
cross}$, for $m^2/(C \Lambda_{\rm cross}^2) = - .5$. The potential
has a runaway form down to the cutoff of the theory} \label{fig1}
\end{figure}

\begin{figure}
\begin{center}
\epsfig{file=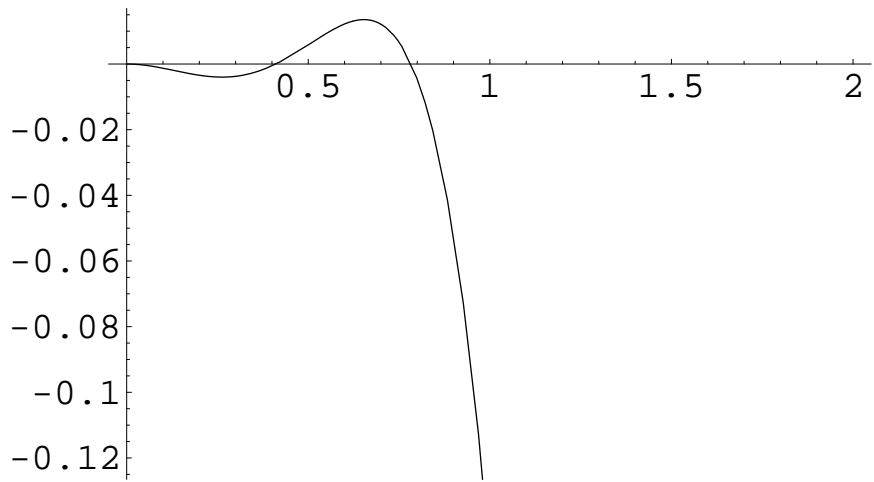,width=0.7\textwidth}
\end{center}
\caption{The effective potential for $m^2/(C \Lambda_{\rm cross}^2)
= - .15$, note the presence of the metastable minimum.} \label{fig1}
\end{figure}

So, there are three possibilities for the gross large scale
structure of the universe in this model: (1) $m^2 > 0$ and no
structure at all, (2) $m^2 < - \Lambda_{\rm cross}^2$ and everything
crumpled near the cutoff close to the Planck scale, or  (3), $m^2 < 0$
but with $|m^2|$ smaller than the scale $\Lambda_{\rm cross}^2$,
where there can finally be a stable minimum with massive particles
lighter than the Planck scale that can make structure. This is very
appealing, and has the same character as Weinberg's argument for the
CC. The {\it only} possibility for structure in this universe occurs
at an exponentially small scale $\Lambda_{\rm cross}$ relative to
the Planck scale. This gives us a new landscape mechanism for
explaining the exponentially large ``weak"-Planck hierarchy in this
model.

By the principle of living dangerously, it must be that $m^2$ is not
{\it too} much smaller than $\Lambda_{\rm cross}$; perhaps no more
than a decade or so lower in energy. This in turn makes a sharp
prediction for the mass of the Higgs particle, since the Higgs
quartic starts at zero near $\Lambda_{\rm cross}$ and only has about
a decade of energy to run up. Therefore the Higgs would be predicted
to be light in this toy universe.

We can extend the lesson from this toy world to build a realistic
model for electroweak symmetry breaking in {\it our} world. As in
the toy world, we will couple the Higgs to new fermions which only
get mass from EWSB, with a neutral one giving dark matter. A minimal
way of doing this is to introduce a vector-like pair of leptons
$(L,E^c), (L^c,E)$ together with two singlets $s,s^c$,  with a
chiral symmetry forbidding all invariant mass terms, and general
Yukawa couplings to the Higgs. After EWSB, all these fermions become
massive near the weak scale, and the lightest particle can be a
neutral Dirac state; for appropriate couplings we can suppress the
direct coupling of this state to the $Z$ to avoid the stringent
direct detection limits on its mass.  Since these particles only get
their mass from EWSB, they will make unsuppressed contributions to
precision EW observables, but with this small particle content the
corrections to $S,T$ can be observably small.

Again, the principle of living dangerously suggests that the Higgs
should be so light that the Higgs quartic coupling is driven
negative less than a decade above the weak scale. For the top mass
near the upper range of its experimentally allowed range $\sim 183$
GeV, and a Higgs mass of $115$ GeV, the Higgs quartic is indeed
driven to zero at about $\sim 10-100$ TeV. This is still
uncomfortably high relative to the weak scale. However the new
Yukawa couplings of the Higgs to the new fermions can make this
happen even more quickly. But it is clear that the Higgs must be
right around the corner in this model. Furthermore, as we have seen,
in order for the vacuum to be sufficiently long-lived, the value of
the Higgs quartic coupling can not become too negative in the
UV--this can be controlled in a number of ways. For instance, new
colored particles can slow-down the running of $\alpha_3$, which in
turn makes $\lambda_t$ smaller in the UV and stops $\lambda_H$ from
becoming too negative at high scales--these particles need not show
up right around the weak scale, though.

In our toy model, for $m^2 > 0$, the Dark Matter particles were
massless and no structure ever formed. In the real world, even if
$m_h^2
> 0$, we still do have EWSB by QCD, and so
the DM particles (and the other fermions) will not be massless,
though they will be exceedingly light. However, as we have discussed
above, the baryon number is drastically suppressed for $m_h^2
> 0$ relative to $m_h^2 < 0$ universes, so there will be no baryons
to make clumping structure for $m_h^2 > 0$.

We leave a detailed construction and analysis of such a model to
future work. However, it is clear that this sort of theory makes a
striking prediction for weak scale physics: a light Higgs must be
discovered, together with new fermions which only acquire mass from
electroweak symmetry breaking, including a neutral state to serve as
dark matter. Once the spectrum and interaction are measured, the
Higgs potential can be reconstructed, and we should find that this
potential is on the edge instability if $m_H^2$ is made slightly
more negative.

\subsubsection{Hierarchy from Baryogenesis}

One of the interesting features of the sort of landscape model
building we are exploring is that aspects of physics beyond the
standard model that are not normally thought to be of fundamental
significance take on elevated importance, if they have to do with
gross infrared properties of the universe. An example is
baryogenesis, since as we have mentioned, baryons are needed to make
interesting clumped structures rather than virialized balls of dark
matter particles.

Since having a non-zero baryon asymmetry is crucial for generating
structure, we want to explore theories where the baryon asymmetry
can {\it only} be generated if the Higgs mass parameters (and
perhaps other scalar masses) are finely tuned to exponentially small
scales. There are a number of mechanisms for producing a Baryon
asymmetry at high scales, most popularly recently from leptogenesis.
Suppose, however, that we are in a neighborhood of the landscape
where none
of these high-energy mechanisms are operative. As we will see in the
theory we construct, there is gauge coupling unification at $\sim
10^{13}$ GeV, which we can take to be the UV cutoff of the theory,
so it is reasonable if the inflationary and reheating scales are
beneath this fundamental scale, that the usual high energy
mechanisms for lepto/baryo genesis are inoperative.

We can still get the required baryon number violation from the
electroweak interactions, but as is well-known, there are two
difficulties faced by models of electroweak baryogenesis in the
Standard Model (see \cite{ckn} for a review). First, there is not
nearly enough CP violation in the Standard Model, as it is
suppressed by the Jarlskog invariant, and second, the electroweak
phase transition in the Standard Model is not sufficiently
first-order, so that the generated baryon asymmetry is washed out
inside the bubbles of broken vacuum.

In order to deal with the latter problem, let us add a singlet
scalar field $S$ to the Standard Model. The scalar potential is of
the form
\begin{equation}
V = \lambda S^4 + \lambda_H (h^\dagger h - \tilde{\lambda} S^2)^2 +
m_S^2 S^2 + m_h^2 h^\dagger h
\end{equation}
where we have assumed an $S \to -S$ symmetry. We will also Yukawa
couple $S$ to fermions $\Psi,\Psi^c$ which are charged under a
non-Abelian gauge group through the interaction
\begin{equation}
\kappa S \Psi \Psi^c
\end{equation}
This continues to respect the $S \to -S$ symmetry with $\Psi,\Psi^c$
having charges $i$. We will assume that this new sector has
confinement and chiral symmetry breaking at its QCD scale determined
by dimensional transmutation
\begin{equation}
\langle \Psi \Psi^c \rangle \sim \Lambda_{{\rm strong}}^3
\end{equation}
which is naturally exponentially smaller than the cutoff of the
theory. We will also assume that this phase transition is
first-order, proceeding via bubble nucleation.

Note that the condensate breaks the $Z_2$ symmetry spontaneously,
leading to a possible domain wall problem cosmologically. However,
this symmetry can also be broken by a small amount explicitly, for
instance by small fermion mass terms for $\Psi,\Psi^c$, so the
domain walls are not an issue.

Let us now examine how the physics varies as we change the
dimensionful parameters $m_S^2,m_H^2$. If $m_H^2$ and $m_S^2$ are
both far larger in magnitude than $\Lambda_{{\rm strong}}^2$, then
the interaction with the new strong sector is completely irrelevant.
The new interactions involving $S$ do not help make a more
first-order phase transition and the universe is empty of baryons.
Only if $m_S^2$ is comparable to $\Lambda_{{\rm strong}}^2$, is the
dynamics of $S$ affected by the first-order phase transition in the
strong sector at the temperature $\sim \Lambda_{{\rm strong}}$:
inside the bubble of true vacuum, there is effectively a linear term
$\sim \Lambda_{{\rm strong}}^3 S$ which can force a vev for $S$ of
$\sim \Lambda_{{\rm strong}}$ in the bubble. Only if we further have
$m_H^2 \sim \Lambda_{{\rm strong}}^2$, can the $S$ vev then force a
large enough Higgs vev inside the bubble to have $v(T_c)/T_c \gsim
1$ and a sufficiently strong first order electroweak phase
transition.

We have thus found another mechanism to explain exponentially large
hierarchies. The scale $\Lambda_{{\rm strong}}$, is determined by
dimensional transmutation. As the elementary scalar masses $m_S^2$
and $m_H^2$ vary in the landscape, the universe is devoid of baryons
and therefore interesting structure, unless $m_S^2,m_H^2$ happen to
both be finely adjusted to be around the exponentially small scale
$\Lambda^2_{{\rm strong}}$.

\begin{figure}
\begin{center}
\epsfig{file=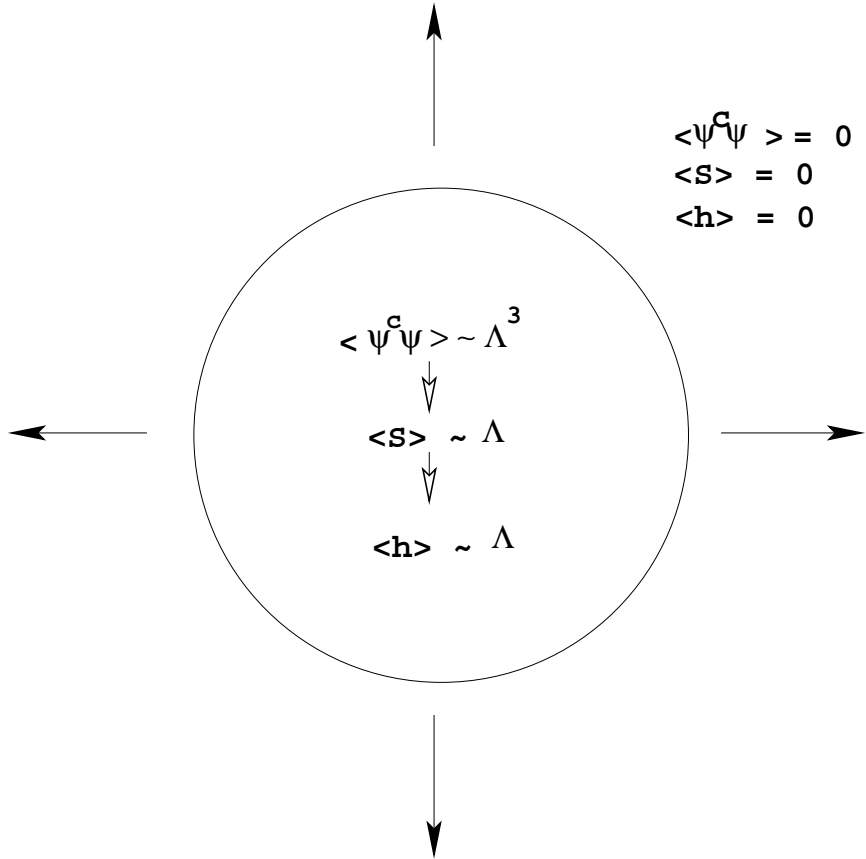,width=0.7\textwidth}
\end{center}
\caption{The first order chiral phase transition in the $\psi$
sector triggers a linear term for the singlet $S$, which in turn
forces a vev for the Higgs field $h$. This then forces a first order
electroweak phase transition}
\end{figure}

In order to actually generate the baryon asymmetry, we need to have
a new source of CP violation as well. The simplest way to do this is
to Yukawa couple $S,h$ to new fermions charged under $SU(2) \times
U(1)$, with new CP violating phases. These fermions can further
serve as the dark matter, if they only get their masses at the weak
scale from the $S,h$ vevs. The simplest possibility is to introduce
a new doublets $\psi_{\pm}$ with hypercharge $\pm 1/2$, as well a
singlet $s$, and Yukawa couplings
\begin{equation}
\kappa S ss + \kappa^\prime S \psi_+ \psi_- + g h^\dagger \psi_+ s +
g^\prime h \psi_- s
\end{equation}
Note that after $S,h$ get vevs, all these fermions become massive,
with a mass near the weak scale. The mass terms for the fermions can
be prohibited by a discrete symmetry under which $h$ is neutral, $S$
flips sign, $s$ has charge $i$ and $\psi_{\pm}$ have charge $-i$.
Note also that there is a single new CP violating phase $\theta$,
associated with the rephase invariant quantity
\begin{equation}
\theta = \mbox{arg} J = \mbox{arg} (\kappa \kappa^\prime g^*
g^{\prime *})
\end{equation}

This particle content is the minimal one that can give us the
required ingredients for electroweak baryogenesis (see \cite{carena}
for a recent attempt to use this sort of particle content for EW
baryogenesis in a different way); small extensions might include
extra singlet fermions. The lightest neutral fermion above is stable
and can be a standard cold dark matter candidate, with a mass
naturally predicted to be right at the weak scale.

Note also that the SM charged particle content of this theory is
precisely the one that we have discussed in the previous subsection,
with only ``Higgsinos" at the TeV scale; therefore, in this model
the couplings unify to good accuracy near $\sim 10^{14}$ GeV.

We leave a full construction of the model and analysis of the
generated baryon asymmetry and dark matter abundance to future work.
But some broad phenomenological consequences are clear. At the LHC,
we expect to produce the new fermions of the theory, as well as the
Higgs, which will have an ${\cal O}(1)$ admixture of the singlet
$S$. We can also make the strongly interacting particles of the
$\Psi$ sector through their Yukawa couplings to the Higgs (via the
$S$ admixture). Furthermore, the model is minimal enough that, once
the appropriate couplings are measured, it is conceivable to
determine whether the baryon asymmetry is explained in the way we
have described. Then, we should once again find that the parameters
$m_h^2, m_S^2$ are near a dangerous edge, such that small variations
in their masses would leave the universe empty of baryons.

Summarizing, this minimal model provides a solution to the CC and
hierarchy problems via the environmental requirement of baryonic
structure, generates the baryon asymmetry, predicts weak-scale Dark
Matter, and has gauge coupling unification. Like all the
landscape-motivated models we are discussing, it also trivially
resolves the little hierarchy problem, since the only new particles
charged under the Standard Model are fermions. As in the case of
split SUSY \cite{Splittwo}, at two-loops, this theory will induce an
electric dipole moment for the SM fermions, which is at the edge of
current experiments for large $\theta$. Note that here, there is a
reason to expect this single phase to be large, as it controls the
CP violation needed for baryogenesis.

\subsection{Other model-building issues}

We have mostly focused on the CC and hierarchy problems, but there
are a host of other aspects of physics beyond the standard model
that need to be re-thought in the context of the landscape. These
include inflation and the generation of density perturbations, the
strong CP problem, the fermion mass hierarchy, neutrino physics and
so on. It is certainly possible that, in a friendly neighborhood,
there are traditional mechanisms which address all of these puzzles:
perhaps there is a natural theory of inflation, an axion for the
strong CP problem, flavor symmetries or separation in extra
dimensions to explain the fermion mass hierarchy etc, and that none
of the parameters relevant for these solutions are effectively
scanned.

It is however be interesting to explore whether the landscape offers
any qualitatively new possibilities for these problems. Inflation in
particular seems ideally suited to use the landscape--as a number of
authors have argued, inflation may not be natural but nonetheless
occur ``accidentally" somewhere on the landscape
\cite{KKLMMT,NewOld,Tegmark}.

\subsubsection{Environmental first-generation quark masses}

The spectrum of fermion masses and mixings seem to exhibit
interesting patterns; for instance the CKM and mass eigenvalue
hierarchies seem to be related via relations of the form $V_{ij}
\sim \sqrt{m_i/m_j}$, suggesting possible underlying textures for
the Yukawa matrices, which may either reflect some underlying
dynamics via flavor symmetries \cite{FN}, or be a consequence of
separation between fields in extra dimensions \cite{shining, AS}.
Unfortunately, no particularly compelling model of this sort has
emerged. There have also been successes in relating heavy generation
Yukawa couplings via grand unified symmetries, such as $b/\tau$
unification\cite{btau}. There are fewer such successes for the
lightest generation, and there are peculiarities, for instance, the
up-down hierarchy is reversed for the first generation. On the other
hand, the masses of the first generation fermions are the only ones
that can conceivably be environmentally determined, as they are the
most important in determining gross infrared properties of the
universe, such as the existence and structure of atoms. Indeed, the
down quark being heavier than the up quark is crucial for the
existence of hydrogen. So it is interesting to consider the
possibility that the first generation Yukawa couplings scan in the
landscape and are environmentally determined.

There are many ways one might do this. For instance, in a
supersymmetric theory, it can be natural for the landscape chiral
fields $\phi_i$ to have vevs much smaller than $M_*$. If they are
further charged under some flavor symmetry group, the light Yukawa
couplings can have a form
\begin{equation}
\sum_i c_i \left(\frac{\phi_i}{M_*} \right)^n f \, f^c \, H
\end{equation}
If the charges are such that $n$ are odd, the Yukawas will be
scanned around $0$ in the landscape, in a region of width $\sim
(\phi/M_*)^n$.

It is also possible that the $\phi$ vevs are of order the cutoff
$M_*$, and the smallness of the Yukawa couplings is explained by
``Planck slop" after GUT breaking, for instance if the Yukawa
couplings are of the form \cite{slop}:
\begin{equation}
c(\frac{\phi}{M_*})\,\left(\frac{\langle \Sigma \rangle}{M_{{\rm
Pl}}}\right)^n\, f\, f^c \, H
\end{equation}
where, $\langle \Sigma \rangle$ is GUT-breaking vev $\sim 10^{16}$
GeV, and $c(\phi)$ again scans around the origin. In this case, a
range of Yukawa couplings with absolute value smaller than $\sim
(M_G/M_{{\rm Pl}})^n$ is finely scanned.

In both of these examples, the Yukawas uniformly scan a region
around zero.  If the light Yukawa couplings are not protected by
approximate flavor symmetries, but arise from separation in extra
dimensions, either through the ``shining" of flavor breaking from
distant branes \cite{shining} or the localization of the Standard
Model fermions to different fixed points in some dimensions
moderately larger than the cutoff scale \cite{AS}, the distance
between the shining branes or fixed points might be scanned in the
landscape. Since the smallest Yukawa couplings are exponentially
small in these distances, even small variations in the distances can
lead to sizeable variations in the smallest Yukawa couplings, but
over a range between a non-zero minimum and maximum size.

Of course, if the lightest generation Yukawa couplings are scanned,
the atomic principle solution of the the hierarchy problem is lost,
as for much larger Higgs vevs, much smaller Yukawas can be used to
keep the down and up quark masses fixed. As we have shown above,
there can be other environmental solutions of the hierarchy problem.
But it may also be that the light Yukawas scan a region not
including zero, so that arbitrarily larger Higgs vevs are not
allowed, and the atomic principle can still play a role in
explaining the hierarchy.

\section{Outlook}

The strategy for building models on the landscape is very different
from that of the canonical few-vacua theories. In the latter, a
basic principle such as naturalness or unification leads to specific
models and predictivity typically follows from the economy in the
structure of these theories implied by symmetries. Predictivity on
the landscape does not have to come from symmetries. It can come
from environmental or statistical arguments. Environmental arguments
simply follow from the existence of some fragile infrared features
in the universe, such as gravitationally clumped structures in the
form of galaxies. Their existence (``the structure principle") is so
sensitive to the parameters of the theory that it leads to
predictions, such as a small cosmological constant---at least in a
friendly neighborhood of the landscape where the theory is not too
far from the standard model. This argument is analogous to computing
the earth-sun distance from the observation that there is water on
the earth, a fragile feature of our planet. Again, the argument only
works if there are lots of solar systems that are not too different
from our own.

Another example of the same strategy is the ``atomic principle". The
fragile entities in this case are atoms, whose existence fixes the
weak scale---at least in neighborhoods near the standard model, and
where only the weak scale and the vacuum energy are variable. This
principle, together with unification and the hypothesis of
supersymmetry at high energies, motivated the proposal of split
supersymmetry. A new application of the atomic principle is the
minimal model proposed in section 3.2. It is the theory with the
smallest particle content that shares the two main successes of the
SSM, unification and DM.

Another application is the theory of section 3.2.1. There the
fragile feature is the stability of the vacuum and, in the
neighborhood of theories with negative quartic coupling near the
cutoff, it leads to a specific extension of the SM with experimental
consequences. In particular, the statement that the universe lives
dangerously close to metastability can be tested at the LHC. Still
another application is the model of section 3.2.2. There the fragile
feature is the baryon excess, provided we live in a neighborhood
where high-energy baryogenesis is impossible. The requirement of
low-energy electroweak baryogenesis again leads to specific models
with observable consequences. Both of these models predict that the
Dark Matter particle is at the weak scale, and the minimal
implementation of the second model has accurate gauge coupling
unification near $10^{14}$ GeV.

These theories illustrate the point that predictivity in the
landscape results from a combination of  a fragile principle,
together with the hypothesis that we are in a friendly neighborhood
of the landscape where it is hard to satisfy this principle.
Ideally, if we had full control of cosmological evolution in the
landscape, we might  predict which neighborhood of the landscape we
land in and justify the hypothesis that we are in a friendly
neighborhood. In reality, and for sure in practice,  this may not be
possible and, as is often the case, we instead have to rely on
experiment to infer what type of neighborhood we live in. So, to
account for the smallness of the vacuum energy we were led in
section 1 to a friendly neighborhood where only super-renormalizable
terms are effectively scanned. A big bonus of living in such a
neighborhood is that the renormalizable couplings are essentially
unaffected; so it is not an accident that the approach of using
symmetries to relate dimensionless couplings has had some success.

We have given field theory examples of such predictive landscapes in
section 2.  Indeed, if such landscapes can be derived from a
fundamental underlying theory, predictivity is essentially fully
regained. Dimensionless couplings can all be predicted to within a
narrow statistical range, while dimensionful parameters are finely
scanned, but are fixed to hierarchical values by simple
environmental arguments.

In three years, the LHC will begin to reveal the physics of the TeV
scale. For most of the past two decades, it has been anticipated
that colliders would reveal a natural theory for the weak scale,
despite the absence of analogous physics for the cosmological
constant, and the growing unease over the lack of indirect evidence
for new TeV scale particles. We have outlined a new framework for
physics beyond the standard model in this paper, which addresses the
cosmological constant problem simultaneously with the hierarchy
problem, and also explains why we have not yet seen any deviations
from Standard Model predictions in particle physics experiments. The
landscape provides the first set of ideas that allows us to address
all fine-tuning problems on the same footing.  Therefore, while our
understanding of the landscape is still primitive, and it may turn
out to be intractable (or even nonexistent!), it deserves to be
taken seriously.

Like other broad frameworks for model-building, ours does not lead
uniquely to a {\it single} theory and no others. This is also true
for the most canonical framework of four-dimensional gauge theories.
After all, a great realization leading to the construction of the
standard model was that gauge theories should be used to describe
nature, but this in itself did not fix the theory; rather, together
with other hints from data, the Standard Model was constructed as a
concrete example of such a theory, making specific predictions that
could be tested. Similarly, the framework of effective field
theories with extra dimensions is even broader, but has again
suggested specific models that can be confirmed or excluded by
experiment. The same is true of the framework for landscape
model-building we are proposing. Indeed, the models it suggests are
rather rigidly constrained, and make a qualitatively different
character of predictions for LHC physics than the usual natural
theories. Instead of finding a large spectrum of new particles and
interactions typically needed for naturalness, we predict sparse
models with few new particles and couplings, with dimensionful
parameters finely tuned but close to dangerous environmental edges.

\vspace{0.6in} \centerline{\bf{Acknowledgements}} \vspace{0.2in} We
thank H.-C. Cheng, A. Cohen, M. Dine, M. Douglas, G. Dvali, G.
Giudice, J. McGreevy, L. Motl, L. Randall, L. Senatore, E.
Silverstein, A. Strominger, R. Sundrum, L. Susskind, S. Thomas and
L.-T. Wang for useful discussions. Special thanks M. Zaldariaga for
many important cosmological insights. N. A-.H. would also like to
especially thank T. Banks, M. Luty, C. Vafa and S. Weinberg for many
stimulating comments on the landscape, A. Strumia for conversations
suggesting that a theory with only Higgsinos would accurately unify,
and G. Giudice for explaining aspects of Higgsino Dark Matter
phenomenology. The research of N.-A.H. and S.K. was supported by
David and Lucille Packard Foundation Fellowships for Science and
Engineering. S.K. is also supported by the D.O.E. under contract
DE-AC02-76SF00515, and the National Science Foundation grant
0244728. SD is supported by the NSF grant 0244728. N.A-H. is also
supported by the DOE under contract DE-FG02-91ER40654.

\appendix

\section{No Baryons for $m_h^2 > 0$}

In this appendix we flesh out the argument that for a universe with
$m_h^2
> 0$, any baryon asymmetry present in the early universe is washed
out by the electroweak interactions, providing an environmental
argument that rules out this region of parameter space in a theory
scanning $m_h^2$.

Let us see how this happens in a little more detail. The rate per
unit volume for the sphaleron transitions that change the EW
winding number by one unit is given by
\begin{equation}
\Gamma = \alpha_W^4 T (\frac{M_W(T)}{\alpha_W T})^7
e^{-\frac{M_W(T)}{\alpha_W T}}
\end{equation}
where at zero temperature $M_W$ is
\begin{equation}
M_W \sim g_W f \sim g_W \frac{\Lambda_{QCD}}{4 \pi}
\end{equation}

By the anomaly, this transition is accompanied by a unit change in
$B$ and $L$ numbers. However, there is a bias favoring the decrease
in $B$ number: by detailed balance, the ratio of the rates for
increasing vs. decreasing baryon number is
\begin{equation}
\frac{\Gamma_+}{\Gamma_-} = e^{-\frac{\Delta F}{T}}
\end{equation}
where $\Delta F$ is the difference in free energy between the states
with $\Delta B = \pm 1$. In the usual story, $T$ is large relative
to the nucleon masses, so $\Delta F/T \ll 1$, and we have (see e.g.
\cite{ckn} for a nice review)
\begin{equation}
\frac{d n_B}{dt} = \Gamma_+ - \Gamma_- \sim  - \Gamma \frac{\Delta
F}{T}
\end{equation}
In our case, however, $\Delta F/T \sim \Lambda_{QCD}/T$ which can be
much bigger than one, so there is an ${\cal O}(1)$ bias for reducing
$B$, and the rate for B violation is simply given by $\Gamma$
\begin{equation}
\frac{d n}{d t} = -\Gamma
\end{equation}
Even at temperatures moderately beneath $\Lambda$, this rate is so
large that it rapidly drives $n_B$ to zero where the biasing stops.
Actually, $n_B$ is not quite driven to zero, since at some point
$n_B$ gets so low that the annihilation of nucleons and
anti-nucleons freezes out.  This happens for $n_B/T^3 = \eta_*$ where
\begin{equation}
\eta_* \sim \frac{n_*}{s} \sim \frac{\Lambda_{QCD}}{M_{Pl}} \sim
10^{-18}
\end{equation}
This is just the freeze-out abundance for baryon number in a
baryon-symmetric universe.
\section{Statistics Redux}

In this appendix, we review,  for the statistically challenged
reader, the argument leading to the central limit theorem and the
Gaussian distributions we use extensively in this paper.

As in the text, given a quantity $F$ in the vacuum $\{\eta_1,
\cdots, \eta_N \}$ given by
\begin{equation}
F_{\{\eta\}} = \sum_{i = 1}^N \eta_i f_i
\end{equation}
We would like to investigate the behavior of the distribution
\begin{equation}
\rho(F) = \sum_{\{\eta\}} \delta(F - F_{\{\eta\}})
\end{equation}
at large $N$. Lets start by imagining that all the $f_j = \bar{f}$
are identical. In this case, despite the large number of vacua, we
only get $N$ different values of $F$. If $k$ of the $\eta$'s are
$+1$ and the remaining $(N-k)$ are $-1$, the value of $F$ is
\begin{equation}
F_k = k \bar{f} - (N - k) \bar{f} = (2k - N) \bar{f}
\end{equation}
and there are $\pmatrix{N \cr k}$ such vacua, so that
\begin{equation}
\rho (F) = \sum_{k = 0}^N \pmatrix{N \cr k} \delta (F - F_k)
\end{equation}
If we are to be able to efficiently ``scan" for different values of
$F$,  we must therefore have some spread in the different values of
$f_j$. It is clear that for a small spread, the delta functions for
$\rho$ centered around $F_k$ will be broadened into some shape, and
for still larger spread, they will overlap into a single smooth
distribution--we clearly want to be in this limit.

We will now give an approximation to $\rho$  using standard
statistical arguments. Instead of working directly with $\rho$
associated with a specific set of the $f_j$, we imagine that the
$f_j$ have been independently and randomly selected from some
distribution $P(f)$, with
\begin{equation}
\int df P(f) = 1
\end{equation}
As we will see, the details of this distribution won't matter; only
the mean $\bar{f}$ and variance $\Delta f$ will enter
\begin{equation}
\bar{f} = \int d f P(f), \, \, \Delta f ^2 = \int df (f - \bar{f})^2
P(f)
\end{equation}
and these in turn will be matched to the mean and variance in the
$f_j$ directly as
\begin{equation}
\bar{f} = \frac{1}{N} \sum_j f_j, \, \, \Delta f^2 = \frac{1}{N}
\sum_j (f_j - \bar{f})^2
\end{equation}
Then, we consider the smeared $\bar{\rho}$
\begin{equation}
\bar{\rho}(F) = \int df_j P(f_j) \rho(F)
\end{equation}
Expressing the $\delta$ function in the definition of $\rho$ as a
Fourier integral, we express $\bar{\rho} $ in the form
\begin{equation}
\bar{\rho}  = \sum_{\{\eta\}} \int \frac{dk}{2 \pi} \prod_j
\tilde{P}(k \eta_j) e^{i k F}
\end{equation}
Now, it is easy to see that
\begin{equation}
\sum_{\{\eta\}} \prod_j \tilde{P}(k \eta_j) = \left(\tilde{P}(k) +
\tilde{P}(-k) \right)^N
\end{equation}
So we need to approximate $\tilde{P}(k) + \tilde{P}(-k)$. Given the
mean $\bar{f}$ and variance $\Delta f^2$ of $f$, we can approximate
\begin{equation} \tilde{P}(k) = e^{i k \bar{f}} \left(1 -
\frac{1}{2} \Delta f^2 k^2 + \cdots \right)
\end{equation}
and therefore
\begin{equation}
\bar{\rho}(F) \to \int \frac{d k}{2 \pi} \left(e^{i k \bar{f}} +
e^{-i k \bar{f}} \right)^2 e^{-\frac{N \Delta f^2 k^2}{2}} e^{i k F}
\end{equation}
which can easily be integrated to yield
\begin{equation}
\bar{\rho}(F) = \sum_{k = 0}^N \pmatrix{N \cr k} \delta_{\sqrt{N}
\Delta f} (F - F_k)
\end{equation}
where $\delta_w(x)$ is a Gaussian approximant to a delta function,
of width $w$
\begin{equation}
\delta_w(x) = \frac{1}{\sqrt{2 \pi w^2}} e^{-\frac{x^2}{2 w}}
\end{equation}

Thus we have the expected result: the delta functions that we had in
$\rho$ for identical $f_j$ have broadened into Gaussians of width
$\sqrt{N} \Delta f$. Since the $f_k$ are spaced by $\bar{f}$, when
\begin{equation}
\sqrt{N} \Delta f > \bar{f}
\end{equation}
the Gaussians overlap, and we have a smooth distribution. This
condition quantifies the minimum amount of variation in the $f$'s
that is needed to get a smooth distribution for $\rho$.

In this limit, we can approximate $\bar{\rho}$ by a single Gaussian.
Indeed, at large $N$, using Stirlings approximation for $\pmatrix{N
\cr k}$ and replacing the sum over $k$ by an integral, we find
\begin{equation}
\bar{\rho}(F) \to  2^N \left(2 \pi N f^2 \right)^{-1/2}
e^{-\frac{F^2}{2 N f^2}}
\end{equation}
where
\begin{equation}
f^2 = \bar{f}^2 + \Delta f^2 = \frac{1}{N} \sum_i f_i^2
\end{equation}
which is the familiar result from the central limit theorem. This
result could have immediately been obtained simply by approximating
from the start
\begin{equation}
\frac{\tilde{P}(k) + \tilde{P}(-k)}{2} = 1 - \frac{k^2 f^2}{2} +
\cdots
\end{equation}
in our first integral expression for $\bar{\rho}$. In order to
estimate the size of the corrections at finite $N$,  it suffices to
include the terms of order $k^4$ in the expansion of $\tilde{P}(k) +
\tilde{P}(-k)$, which generates terms of order $F^4/N^3$ in the
exponential. Thus the relative corrections are of $O(1/N^2)$.

\section{Instability of $- \phi^4$ theory}

In this appendix we outline an analysis of the instability in $-
\lambda \phi^4$ theory, which is needed in analyzing the model with
a negative quartic coupling for the Higgs at very high scales; see
e.g. \cite{strumiavac} for an analysis relevant to vacuum stability
in the Standard Model. It is easy to see that any instability of the
perturbative vacuum at $\phi = 0$ is non-perturbatively small in
$|\lambda|$--in fact despite the absence of any barrier in the
potential, the vacuum decay is a tunneling process. An intuitive way
of seeing this is as follows. Suppose we start with the perturbative
vacuum with $\phi = 0$.  Of course the energy can be lowered if we
go further down the potential, but in order to do this in a bubble
of size $R$ also costs some spatial gradient energy. Indeed, if we
consider a bubble where the field has a value $\phi_0$, then to
break even energetically we have to have

\begin{equation}
E = R^3 \left((\frac{\phi_0}{R})^2 - \lambda \phi_0^4 \right) = 0
\rightarrow \phi_0^2 = \frac{1}{\lambda} \frac{1}{R^2}
\end{equation}
Now, what is the amplitude to find the ground state in this
configuration? It is easy to get from the Harmonic oscillator
wavefunction for the ground state
\begin{equation}
\Psi \propto e^{-\int d^3 k \tilde{\phi}(k) |k| \tilde{\phi}(-k)}
\end{equation}
and we find the amplitude is
\begin{equation}
e^{-R^2 \phi_0^2} \sim e^{-1/\lambda}
\end{equation}
This is of course in accord with the fact that the instability in
the vacuum is invisible in perturbation theory.

A proper treatment of the tunneling rate can be done with the
standard Euclidean methods. Despite the absence of a barrier in the
potential, there is indeed a bounce solution to the Euclidean
equations of motion. In fact, due to the classical scale invariance
of the theory, there is a continuum of solutions labeled by the
bubble size $\rho$
\begin{equation}
\phi_{bounce}(r) = \frac{r}{r^2 + \rho^2}
\end{equation}
The bounce action is $\rho$ independent and is given by
\begin{equation}
S = \frac{16 \pi^2}{3 |\lambda|}
\end{equation}
Obviously one-loop corrections break the classical scale invariance,
and to leading order the contribution of a bubble of size $\rho$ to
the tunneling amplitude is the same as the above with the running
$\lambda(1/\rho)$, leading to an expression for the decay rate per
unit volume
\begin{equation}
\frac{\Gamma}{V} = \int \frac{d \rho}{\rho^5} \, e^{-\frac{16
\pi^2}{3 \lambda(1/\rho)}}
\end{equation}
The lifetime can be long enough as long as the quartic coupling
doesn't get too negative in the UV.

\end{document}